\documentclass[apj,numberedappendix]{emulateapj}

\usepackage{natbib}
\bibpunct{(}{)}{;}{a}{}{,}
\usepackage{color}
\definecolor{darkgreen}{RGB}{0,142,128}

\usepackage{hyperref}
\hypersetup{colorlinks,citecolor=blue,linkcolor=blue}

\usepackage{amsmath}
\usepackage{amsthm}
\usepackage{amsfonts}
\usepackage{url}
\usepackage{subfigure}
\usepackage{multirow}

\newcommand{\modif}[1]{{#1}}
\newcommand{\modd}[1]{{#1}}
\newcommand{\reff}[1]{{#1}}

\begin{document}

\title{On the Diversity of Magnetic Interactions in Close-In
  Star-Planet Systems}
\shorttitle{On the Diversity of Magnetic Interactions in Close-In
  Star-Planet Systems}

\author{A. Strugarek}
\affil{D\'epartement de physique, Universit\'e de Montr\'eal, C.P. 6128 Succ. Centre-Ville, Montr\'eal, QC H3C-3J7, Canada}
\affil{Laboratoire AIM Paris-Saclay, CEA/Irfu Universit\'e Paris-Diderot CNRS/INSU, F-91191 Gif-sur-Yvette.}
\email{strugarek@astro.umontreal.ca}
\author{A. S. Brun}
\affil{Laboratoire AIM Paris-Saclay, CEA/Irfu Universit\'e Paris-Diderot CNRS/INSU, F-91191 Gif-sur-Yvette.}
\author{S. P. Matt}
\affil{Astrophysics group, School of Physics, University of Exeter, Stocker Road, Exeter EX4 4QL, UK }
\author{V. R\'eville}
\affil{Laboratoire AIM Paris-Saclay, CEA/Irfu Universit\'e Paris-Diderot CNRS/INSU, F-91191 Gif-sur-Yvette.}

\shortauthors{Strugarek, et al.}
\begin{abstract}
Magnetic interactions between close-in planets and their host star
can play an important role in the secular orbital evolution of the
planets, as well as the rotational evolution of their host. As long as
the planet orbits inside the Alfv\'en surface of the stellar wind, the
magnetic interaction between the star and the planet can modify the
wind properties and also lead to direct angular momentum transfers
between the two.
We model these star-planet interactions using compressible
magneto-hydrodynamic (MHD) simulations, and quantify the angular
momentum transfers between the star, the planet, and the stellar
wind. We study the cases of magnetized and
non-magnetized planets and vary the orbital radius inside the
Alfv\'en surface of the stellar wind. Based on a grid of numerical
simulations, we propose general scaling laws 
for the modification of the stellar wind torque, for the torque
between the star and the planet, and for the planet migration
associated with the star-planet magnetic interactions. We show that
when the coronal magnetic field is large enough and the star is
rotating sufficiently slowly, the effect of the magnetic star-planet
interaction is comparable to tidal effects and can lead to a rapid
orbital decay. 
\end{abstract}


\maketitle

\section{Introduction}
\label{sec:introduction}

\reff{More than a thousand planets have now been discovered orbiting
  distant stars. These} planets span several orders of magnitudes in mass, radius and
\modif{semi-major axis, and $187$} of them \modd{to date} orbit very
close  ($r_{orb} < 10\, r_{\star}$) to their
host\footnote{\url{http://exoplanet.eu/}}. Due to their proximity they are
interacting with their star in very
different physical conditions in terms of
interplanetary plasma density, pressure, wind velocity, and magnetic field
strength, compared to any other planet in our solar system. 
Star-planet interactions (SPIs) can originate from tidal forces, 
magnetic fields, winds, and radiative processes \citep[see ][]{Cuntz:2000ef}. They
have local and global consequences on the system over a large range of time-scales. 

SPIs may cause
enhanced chromospheric and coronal activity. For instance,
evidence of chromospheric hotspots related to an orbiting planet have been
observed in several systems \citep{Shkolnik:2005bz} and were
theoretically modeled by \citet{Lanza:2008fn,Lanza:2012jv}. For
massive and sufficiently close planets, it was suggested that SPIs
could lead to an overall increase of the stellar magnetic activity
\reff{\citep[\textit{e.g.}, traced by an increase in the X-ray and UV emissions from the
star, see][]{Kashyap:2008gi,Shkolnik:2013aa}} due
to tidal \citep{Cuntz:2000ef} or magnetized \citep{Cohen:2011gg}
interactions. In addition, SPIs were
also proposed to be at the origin of
super-flares \citep{Rubenstein:2000hp}, although the lack of correlations
between super-flaring stars and hot-Jupiter hosts observed with Kepler \citep{Shibayama:2013ji}
suggests other triggering mechanisms
\citep[see][]{Shibata:2013fo}. \reff{Magnetic dynamos operate in the
  interior of stars and planets \citep[for recent reviews,
  see][]{Stevenson:2003ji,2010LRSP....7....3C,Jones:2011fx,Brun:2013kc}.}
  It was recently suggested that even the
  dynamo operational mode itself may be influenced by the
  star-planet (tidal) interactions
  \citep{Abreu:2012fu,Charbonneau:2013aa}, leading to potentially observable
  perturbations of, \textit{e.g.}, the spot cycle of the Sun.

 On the planetary side,
\citet{Zarka:2007fo} proposed that the magnetized SPIs (SPMIs) could
lead to enhanced radio emissions in the planetary
magnetosphere. \citet{Jardine:2008ec} characterized such emissions and
showed they mainly depend on the density and magnetic field profiles in
the stellar wind \citep[e.g., see][for a detailed theoretical
modelling of radio emissions in the $\tau$ Boo
system]{Vidotto:2012ks}. \modd{Enhanced evaporation of the planetary
atmosphere, due to stellar coronal activity, has also been reported by
\citet{Lecavelier:2010kw}. Although
the various emission enhancements
have not been systematically observed in close-in planet systems
\citep{Donati:2008hw,Fares:2010hq}, \citet{Scharf:2010ab} showed that
the observed positive correlation between the X-ray luminosity of the
system and the mass of the orbiting planet could be used as a probe to
measure the planetary magnetic field.} Hence, the various emission enhancements that may originate
from SPIs could be used, at least in principle, to estimate some
physical properties of exoplanets. The temporal
variability, as well as the physical mechanisms at the origin of those
emissions, are still today an active subject of research \citep{Shkolnik:2008gw,Miller:2012gq}. 

SPIs also have a major influence on the global properties of star-planet systems. For
instance, understanding the stellar radiation and the stellar wind local properties
is key to
determine how a planet interacts with its environment \citep[see,
e.g,][]{Lammer:2009im}, and
ultimately to determine the zone of habitability 
around stars \citep{Selsis:2007ds}. 
Tidal interactions are well known to lead to the \modif{spin-}orbital
synchronization \modif{(\textit{e.g.} through the so-called
  tidal-locking mechanism)} of close-in planets. They also have more subtle effects in
star-planet systems \citep[for a review, see][]{Mathis:2013cd} and can
for instance affect
the orbital evolution of the planet \citep{Bolmont:2012go,AuclairDesrotour:2014io,Zhang:2014iz}
or even the stellar rotational evolution
\citep[][]{Barker:2011jn,Poppenhaeger:2014be}. In addition, 
magnetic interactions
result in a torque applying to the
orbiting planet, which also influences its migration
\citep{Laine:2008dx,Lovelace:2008bl,Vidotto:2010iv,Lanza:2010bo,Laine:2011jt}. If the
planet orbits \modif{inside the Alfv\'en surface of the stellar wind
  (defined as the surface where the wind speed equals the local
  Alfv\'en speed)}, torques apply to
the star as well and lead to a modification of its rotational history
\citep{Cohen:2010jm}. In some extreme cases, the SPIs can lead to
the expansion of the planetary atmosphere beyond its Roche
lobe, resulting in a constant outflow from the planet to the star
which will also affect the orbital properties, as well as the
stellar rotational history \citep{Lai:2010he}.

The development of a model describing the numerous star-planet interactions,
for the different types of stars and planets,
is a formidable challenge but is extremely valuable
for our understanding of the creation and evolution of planetary systems, and for
the characterization of the observed exoplanetary systems. \modif{An ultimate
goal is to} develop a theoretical framework, based on numerical simulations, in
which all the SPI effects could be taken into
account, self-consistently. We focus the present work on the less-studied
magnetized interaction, and more specifically on aspects of long-term
impacts a close-in planet can exert on its host star. \modif{This simplified
model uses a 2.5D (axisymmetric) geometry for simplicity, and will
provide the basis for future and more detailed models.}

Various definitions of the term \textit{close-in} planet have
been used in the literature. Here, we define a close-in planet 
as a planet that is able to influence its host star through magnetic
interactions. Said differently, we consider close-in planets 
to be orbiting inside the Alfv\'en radius of the stellar wind. Alfv\'en waves excited by the
presence of an orbiting planet can then travel from the planet
vicinity to the stellar surface, where they are able to modify
the plasma properties. Most studies of the SPMIs so far
have been focused on their effects on the planetary dynamics \citep[either fast
magnetospheric evolution or slow planet migration, see][for a recent
exemple]{Cohen:2014eb}, rather than describing the
important feedback such a planet can exert on its host star on a
secular time-scale
\citep[see][for a notable first study of such
long-term effect]{Cohen:2010jm}. The long term impact of SPMI can be two-fold:
the magnetic torque leads to a direct transfer of 
angular momentum between the two bodies, and the magnetic interaction
can modify the wind driving in the stellar corona. Modelling the
latter requires taking into account coronal feedbacks in the wind
driving mechanism. 

\modif{We build our study on stellar wind models pioneered by
\citet{Washimi:1993vm} and further developed by, \textit{e.g.},
\citet{Keppens:1999tw,Matt:2004kd,Matt:2008bj,Matt:2012ib,Strugarek:2014fr,Reville:2014ud}.}
These models
possess good
conservation properties and are designed to adapt to external
perturbations \citep{Strugarek:2014fr}. We develop in this work a
numerical model for thermally driven winds (Section 
\ref{sec:model-stell-wind}), in which close-in planets
\modif{are introduced at various orbital radii} (Section
\ref{sec:modeling-planets}). We investigate the cases
of magnetized --with different topologies-- and unmagnetized planets,
to systematically characterize the magnetized angular momentum transfers occurring in star-planet
systems, along with the modification of the stellar wind induced by the SPMI (Section
\ref{sec:magnetic-torques}). We propose scaling
laws for the effect of SPMIs in Section \ref{sec:discussion} and 
summarize our main findings in Section \ref{sec:conclusions}.

\section{Stellar wind model}
\label{sec:model-stell-wind}

We compute solutions for steady-state stellar winds, using the finite
volume magneto-hydrodynamic (MHD) code PLUTO
\citep{Mignone:2007iw}. We detail in Section
\ref{sec:simulation-method} our simulation method and 
in Section \ref{sec:select-part-stell} the fiducial stellar wind model
selected to study the SMPIs.

\subsection{Simulation method}
\label{sec:simulation-method}

The PLUTO code solves the following set of ideal MHD equations:
\begin{eqnarray}
  \label{eq:mass_consrv_pluto}
  \partial_t \rho + \boldsymbol{\nabla}\cdot(\rho \mathbf{v}) &=& 0 \, \\
  \label{eq:mom_consrv_pluto}
  \partial_t\mathbf{v} +
  \mathbf{v}\cdot\boldsymbol{\nabla}\mathbf{v}+\frac{1}{\rho}\boldsymbol{\nabla} P
  +\frac{1}{\rho}\mathbf{B}\times\boldsymbol{\nabla}\times\mathbf{B}
  &=& \mathbf{g} \, ,
  \\
  \label{eq:ener_consrv_pluto}
  \partial_t P +\mathbf{v}\cdot\boldsymbol{\nabla} P + \rho
  c_s^2\boldsymbol{\nabla}\cdot\mathbf{v} &=& 0 \, ,\\
  \label{eq:induction_pluto}
  \partial_t \mathbf{B} - \boldsymbol{\nabla}\times\left(\mathbf{v}\times\mathbf{B}\right)
  &=& 0 \, ,
\end{eqnarray}
where $\rho$ is the plasma density, $\mathbf{v}$ its velocity, $P$ the gas
pressure, $\mathbf{B}$ the magnetic field, $\mathbf{g}$ the
gravitational acceleration \modd{(which is time-independent)}, and
$c_s=\sqrt{\gamma\,P/\rho}$ the sound
speed \reff{($\gamma$ is the adiabatic exponent, taken to be the equal
  to the ratio of specific heats)}. \modif{We use an
ideal gas equation of state
\begin{equation}
  \label{eq:EOS}
  \rho\varepsilon = P/\left(\gamma-1\right)\, ,
\end{equation}
where $\varepsilon$ is the internal energy \reff{per mass.}}

We use the following numerical method implemented in the PLUTO code. 
First, a minmod
limiter is used on all the variables, combined to a \textit{hll} 
(Harten, Lax, Van Leer) solver to compute the intercell fluxes. A
second order Runge-Kutta scheme is used for the time
evolution. The solenoidality of the magnetic field
($\boldsymbol{\nabla}\cdot\mathbf{B}=0$) is ensured with a
\textit{constrained transport} method \citep[see
][]{Evans:1998aa,Gardiner:2005ky}. We use a 2.5D geometry, centered on
the rotating star, \modd{meaning that we solve the equations for fully 3D
vector components of the velocity and magnetic fields, but assume an
axisymmetric geometry}.

Following the work of \citet{Matt:2012ib}, we initialize our
simulations with a spherically symmetric,
hydrodynamic Parker wind solution \citep{Parker:1958dn}, to which we add a dipolar
magnetic field with a magnetic moment $\mu_{\star}$. We developed
special boundary conditions at the base of the wind \citep[after][]{Matt:2004kd} that
ensure good conservation properties \citep{Lovelace:1986kd,
  Keppens:2000ea,Zanni:2009kc} along the magnetic field
lines. \reff{They consist of three circular layers representing the
  lower corona in which the Parker wind pressure gradient, the
rotation rate of the star, and its magnetic field are successively
imposed. They are designed as follows:} 
\reff{\begin{itemize}
\item[-] \textbf{Upper Layer}. The density and pressure are fixed to the (1D) Parker
wind solution. The poloidal $(\varpi,z)$ velocity field is forced to
be parallel to the poloidal magnetic field, while its magnitude can
evolve freely. The azimuthal velocity and
the magnetic field are left free to evolve.
\item[-] \textbf{Middle Layer}. The density and pressure are fixed to the (1D) Parker
wind solution. The poloidal velocity is set to zero, and the azimuthal
velocity is fixed to the stellar rotation. The magnetic field is left
free to evolve.
\item[-] \textbf{Lower Layer}. The density and pressure are fixed to the (1D) Parker
wind solution. The poloidal velocity is set to zero and the azimuthal
velocity is fixed to the stellar rotation. The poloidal magnetic field
is fixed to the dipolar stellar field. In the open field lines regions
$B_{\varphi}$ is set to minimize the poloidal currents. In the closed
field lines regions $B_{\varphi}$ is set to zero. The open and closed
field lines regions are distinguished based on a local criterion involving
the azimuthal Alfv\'en speed, which allows us to dynamically identify
the various regions as the wind evolves.
\end{itemize}}
\reff{We also observed that adding one more layer above the stellar
surface, where the poloidal velocity field is forced to be parallel to 
the poloidal magnetic field, improves the conservation properties of
the model. All the models presented in this work use this
additional layer.}

\reff{When a planet is included in the model, the density and pressure are held fixed to
specified values in the planetary interior. The poloidal velocity
is fixed to zero and the
azimuthal velocity to the keplerian velocity in the whole planetary interior. In the unipolar cases
(see Section \ref{sec:unipolar-interaction}), the magnetic field is
free to evolve inside the planet, and in the 
dipolar cases (see Section \ref{sec:dipolar-interaction}), it is held fixed
to the planetary dipole.}

\reff{Our boundary conditions allow
the wind driving to
automatically adapt to external perturbations originating from the
star-planet magnetic connection. We impose outflow 
conditions on the outer boundaries (zero-gradient imposed on all quantities), and axisymmetry at the rotation
axis. We refer the reader 
to \citet{Strugarek:2014fr} for a more
complete discussion on boundary conditions.}

\begin{figure}[tbp]
  \includegraphics[width=\linewidth]{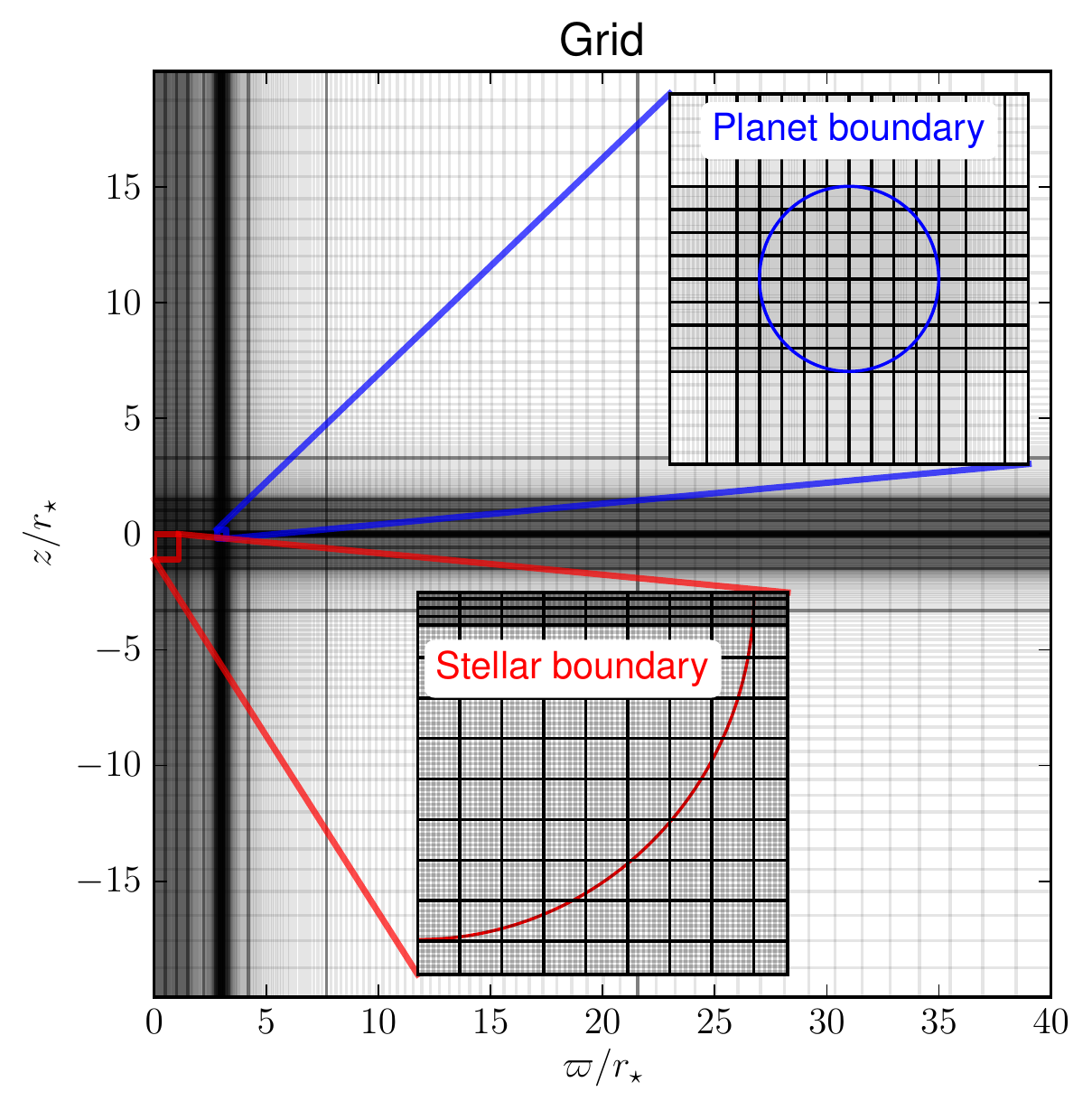}
  \caption{Grid of the $r_{orb}=3\, r_{\star}$ cases. The grid is
    highlighted every 32 points. The two insets
    are zooms on the southern hemisphere of the star and on the planet
  location. The boundaries of the star and the planet are labeled in
  red and blue. In those insets the grid is highlighted every eight points.}
  \label{fig:grid}
\end{figure}

\modif{\modd{The \modif{stellar radius} is uniformly discretized over 64
  grid points in the radial direction and is discretized over two
  uniform domains in the vertical direction to ensure a higher
  resolution on the equatorial plane where the planet lies.} The circular planet is
  discretized with a square of 64$^2$ grid points (we adapt the grid
  for each planetary orbital radius considered in this work). \reff{The rest
  of the domain is discretized with stretched grids,}
  giving a typical overall
  resolution of 400 points in the radial direction and 384 in the
  vertical direction. The grid for models with an orbital
  radius of $3\, r_\star$ is shown in Figure \ref{fig:grid}, with two
  insets zooming on the stellar and planetary boundaries.
}

In the planet-free case, a steady-state
stellar wind is typically obtained after a few sound crossing times, when
the accelerating wind and rotating
magnetic field are dynamically balanced. The solution
for the wind depends on three velocity ratios defined at the surface
of the star, and on the ratio of specific heats $\gamma$. The three
characteristic velocities are the sound speed $c_{s}$,
the Alfv\'en speed $v_A=B_{\star}/\sqrt{4\pi\rho_{\star}}$ (where
$B_{\star}=\mu_{\star}/r_{\star}^{3}$ is the magnetic field strength at the stellar
equator) and the rotation speed 
$v_{\rm rot}$ (in this work, the
star is considered to rotate as a solid body). Their ratios to the escape velocity $v_{\rm
  esc}=\sqrt{2GM_\star/r_\star}$ at the stellar surface then define a
unique stellar wind solution. The \modif{global properties of the} wind can be 
characterized by its mass loss rate
rate $\dot{M}_\star$ and its angular momentum loss rate (AML)
$\dot{J}_\star$, which are computed a posteriori and defined by \citep[\textit{e.g.},][]{Matt:2004kd}
\begin{eqnarray}
  \label{eq:mass_loss_rate}
\dot{M}_{\star} &=& \oint{\rho \mathbf{v} \cdot  d\mathbf{A}} \, , \\
  \label{eq:angmom_flux}
\dot{J}_\star &=& \oint{ \varpi \left( v_\phi -
    B_\phi\frac{\mathbf{v}_p\cdot\mathbf{B}_p}{\rho|\mathbf{v}_p|^2}
  \right) \rho\mathbf{v}\cdot d\mathbf{A}}\, ,
\end{eqnarray}

where $(\varpi,\phi,z)$ is the cylindrical coordinate system, the subscript $p$ denotes the
poloidal $(\varpi,z)$ component of a vector, and
$\oint \mathbf{x}\cdot d\mathbf{A}$ stands for the integral of 
$\mathbf{x}$ on a spherical surface enclosing the star. 
Because mass and momentum are conserved\modd{, a steady-state requires} the spherical
integrals \eqref{eq:mass_loss_rate} and \eqref{eq:angmom_flux} to be
constant in between sources and sinks (here, the star and the planet)
in the domain.
Hence, the integrals can be \modif{equivalently evaluated on any spherical surface when a 
statistical steady-state is reached in the simulation}.

\subsection{Fiducial stellar wind}
\label{sec:select-part-stell}

Depending on the choice of parameters, the simulated stellar winds possess
a variable size dead-zone \citep[\textit{i.e.} the zone where 
the magnetic pressure is high enough to confine the
plasma and suppress the wind driving; see,
e.g.,][]{Mestel:1968aa,Keppens:1999tw,Matt:2008bj}. The relative position
of the Alfv\'en surfaces and the dead-zone radius can also vary
significantly.  The three Alfv\'en surfaces 
label the position in the stellar wind at which the three
following magnetic Mach number are equal to unity \citep{Keppens:1999tw},
\begin{eqnarray}
  \label{eq:v_alfvenic}
  &\left(M_A\right)^2 = \frac{v_\varpi^2+v_z^2+v_{\phi}^{2}}{A_\varpi^2+A_z^2+A_{\phi}^{2}}  \, ,\\
  \label{eq:v_alfvenic_slow}
  &\left(M_s\right)^2 =
  \frac{2\left(v_\varpi^2+v_z^2\right)}{c_s^2+A_p^2+A_\phi^2-
    \sqrt{\left[c_s^2+A_p^2+A_\phi^2\right]^2-4c_s^2A_p^2}}
  \, , \\
  \label{eq:v_alfvenic_fast}
  & \left(M_f\right)^2 =
  \frac{2\left(v_\varpi^2+v_z^2\right)}{c_s^2+A_p^2+A_\phi^2+
    \sqrt{\left[c_s^2+A_p^2+A_\phi^2\right]^2-4c_s^2A_p^2}} \, ,
\end{eqnarray}
where $\mathbf{A}=\mathbf{B}/\sqrt{4\pi\rho}$. We use
$r_{a}$ and $r_{f}$ to denote the positions of the Alfv\'en surface ($M_{A}=1$)
and fast Alfv\'en surface ($M_{f}=1$) on the
equatorial plane. We build a simulation with
parameters listed in Table \ref{tab:tab1}, following on the preliminary
study of \citet{Strugarek:2012th}. Because we work with dimensionless
quantities, a single wind
simulation may represent different physical winds depending on the
normalization. \modd{For a given simulation,} the density normalization
$\rho_{0}$ directly sets the physical amplitude of the stellar magnetic field.
The physical mass loss rate 
\reff{is then determined by the radius and mass of the star, which we
  consider here to be solar}.
\modif{We give in Table \ref{tab:tab1} their physical values, as well as the
mass and angular momentum loss rates of the modelled stellar wind for
two possible density normalizations.}
\modif{The two density normalizations were chosen to
represent a solar-like mass loss rate in the first case, and in the
second case a
very large mass loss rate that is thought to be representative
of young Suns or T Tauri stars. It corresponds to a
variation of five orders of magnitude of $\rho_{0}$. The magnetic
field of $246$ G in the second case is quite large, it is likely to
represent an upper limit case
for the potential effects of the SPMIs. Finally we define the
torque applied by the (fiducial) wind to the star, 
$\tau_{w}=-\dot{J}_\star$, which will be used as a normalization to the
magnetic torques in the remainder of this work.} \reff{In all cases, the modelled
  stellar wind reaches $450$ km/s near $1$~AU, which is representative
  of the 'slow' component of the solar wind.}

\begin{deluxetable}{lc}
  \tablecaption{Stellar wind parameters and characteristics\label{tab:tab1}}
  \tablecolumns{2}
  \tabletypesize{\scriptsize}
  \tablehead{
    \colhead{Parameter} &
    \colhead{Value}
  }
  \startdata 
  $\gamma$ & 1.05 \\
  $c_s/v_{\rm esc}$ & 0.2599 \\
  $v_A/v_{\rm esc}$ & 0.3183 \\
  $v_{\rm rot}/v_{\rm esc}$ & 0.00303 \\
  \hline \hline  \\
  \multicolumn{1}{c}{Characteristics}  & Value\\
  \hline \\
  $\rho_{0}$ [g cm$^{-3}$] & 1.3e-16 | 1.3e-11\\
  $B_{\star}$ [G] & 0.78 | 246\\
  $\dot{M}_\star $ [$M_\odot$ yr$^{-1}$]  & 2e-14 | 2e-9 \\ 
  $\dot{J}_\star$ [$M_\odot R_\odot^2$ yr$^{-2}$] & 4e-11 | 4e-6
  \enddata
\end{deluxetable}

\begin{figure}[tbp]
  \includegraphics[width=\linewidth]{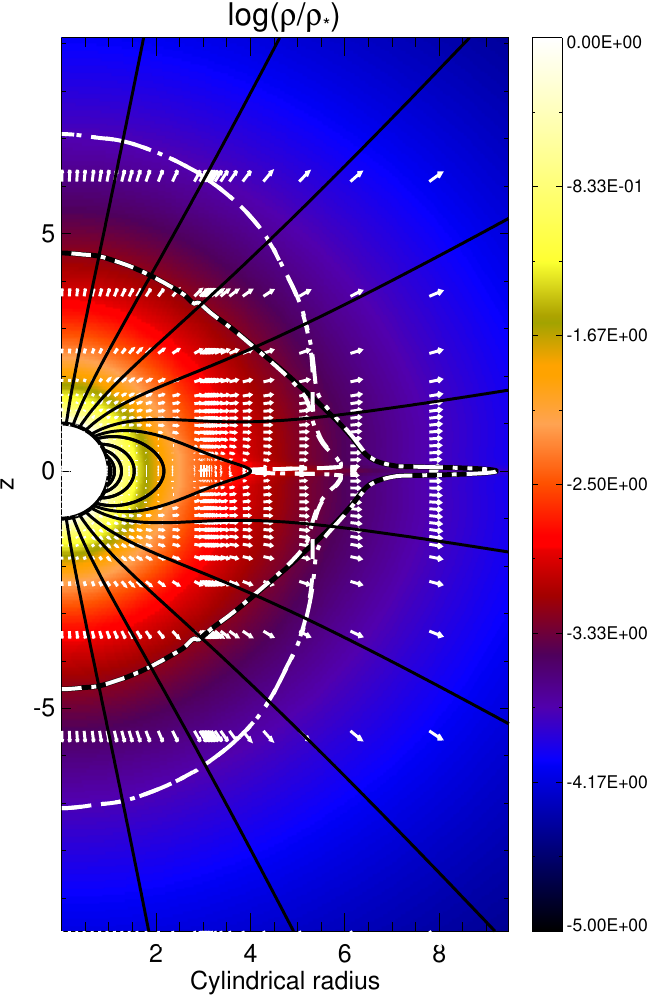}
  \caption{Fiducial stellar wind. Axes are given in units of stellar radius.
    The color map represents the logarithm of the wind density,
    normalized to the stellar surface density. White arrows show the
    local poloidal wind velocity. The magnetic field lines are
    displayed in solid black lines. Four characteristic surfaces are
    shown: the fast (three dot-dash white line), slow (dot-dash
    white line) and classic (dashed white line) Alfv\'en surfaces \reff{(see
    Equations \ref{eq:v_alfvenic}-\ref{eq:v_alfvenic_fast})}, and
    the surface at which the local Mach number is equal to one (thick
    black line). The stellar surface is indicated by a black half circle.}
  \label{fig0}
\end{figure}

The resulting stellar wind is displayed in Figure \ref{fig0}. The
initial magnetic dipole has opened up under the influence of the
accelerating wind, leaving a dead-zone near the equator where
the field lines reamin closed. 
The three Alfv\'en surfaces are delimited by the white lines. 
In this case, the Alfv\'en surface coincides with the fast Alfv\'en surface at the poles
and with the slow Alfv\'en surface ($M_s=1$) at the equator
\citep{Keppens:1999tw}. Note that the surface where the Mach number
is one ($\mathbf{v}_p=c_s$, black line in Figure \ref{fig0}) coincides with those surfaces in an
exactly opposite way. We obtain the position of the
  Alfv\'en surface on the equatorial plane 
  $r_{a} \sim 3.5\,  r_\star$ and the position of the fast
Alfv\'en surface averaged on a thin disk centered on the equator 
$r_{f}\sim 7\, r_\star$.

We focus in this work on the magnetic feedback close-in planets can
exert on their host
stars. Such planets necessarily orbit inside the fast-Alfv\'en surface
to enable Alfv\'en waves to travel from the planet to 
the stellar surface. Indeed, a planet orbiting outside the Alfv\'en
surfaces can be influenced by the magnetized stellar wind
\citep[\textit{e.g.},][]{Vidotto:2014kk}, by stellar radiation and by
tides, but it cannot back-react magnetically on its host star. The
SPMI is driven by the differential motion between the orbiting planet and the
rotating wind. The positions of the Alfv\'en surfaces and the
rotation rate of the wind can be a priori estimated from our knowledge of
thermally driven stellar wind.
On the equatorial plane, for a dipolar-type thermally driven stellar wind, a dead-zone
co-rotating with the star extends roughly from the stellar surface to
the Alfv\'en surface. The fast Alfv\'en surface generally extends 
further away from the star, and the rotation profile of the region
in between \modif{roughly} falls proportionally to the inverse of the distance to the
star. The \modif{approximate} equatorial position of the Alfv\'en surface can be
semi-analytically derived with the same methodology used by
\citet{Matt:2012ib} and \citet{Reville:2014ud}.
\citet{Matt:2012ib} showed
that a stellar wind could be unambiguously characterized, either by its mass-
and angular momentum loss rates, or, equivalently, by the stellar
equatorial rotation speed $f$ and a parameter $\Upsilon$, representing
a dimensionless mass loss rate \citep[similar to the magnetic
confinement parameter of][]{udDoula:2002km} and defined by
\begin{eqnarray}
  \label{eq:def_f}
  f &=&
  \Omega_{\star}r_{\star}^{3/2}\left(GM_{\star}\right)^{-1/2} =
  \sqrt{2}\, \frac{v_{rot}}{v_{esc}}\, ,\\
  \label{eq:def_upsilon}
  \Upsilon &=& B_{\star}^{2}r_{\star}^{2} \left(
    \dot{M}_{\star}v_{esc}\right)^{-1} =
  \frac{4\pi\rho_{\star} r_{\star}^{2} v_{esc}}{\dot{M}_{\star}}\left(\frac{v_{A}}{v_{esc}}\right)^{2}\,
  , 
\end{eqnarray}
where $\Omega_{\star}$ is the stellar angular rotation rate. 
Using the grid of stellar wind models described in \citet{Reville:2014ud}, we fit the
Alfv\'en equatorial radii\footnote{\modif{Note that this is an
  \textit{equatorial} value, and not the torque-determined globally
  averaged value used in \citet{Matt:2012ib,Reville:2014ud}}} to the wind
parameters $(\Upsilon,f)$ with
\begin{eqnarray}
  \label{eq:ra_fit}
  \frac{r_{a,f}}{r_{\star}} =
  K_{1}^{a,f}\left(\frac{\Upsilon}{\sqrt{1+(f/K_{2})^{2}}}\right)^{m^{a,f}}\, .
\end{eqnarray}
The fitted parameters $K_{1}$, $K_{2}$ and $m$ are given in Table \ref{tab:tab0}. 

\begin{deluxetable}{cccc}
  \tablecaption{Fit parameters for the position of Alfv\'en
    surfaces on the equatorial plane. \label{tab:tab0}}
  \tablecolumns{4}
  \tabletypesize{\scriptsize}
  \tablehead{\colhead{} &
    \colhead{$K_{1}$} &
    \colhead{$K_{2}$} &
    \colhead{$m$}
  }
  \startdata 
  $r_{a}$ & $2.3027$ & $0.0014$ & $0.1842$ \\
  $r_{f}$ & $4.8412$ & $0.0027$ & $0.1858$ 
  \enddata
\end{deluxetable}

\begin{figure}[tbp]
  \includegraphics[width=\linewidth]{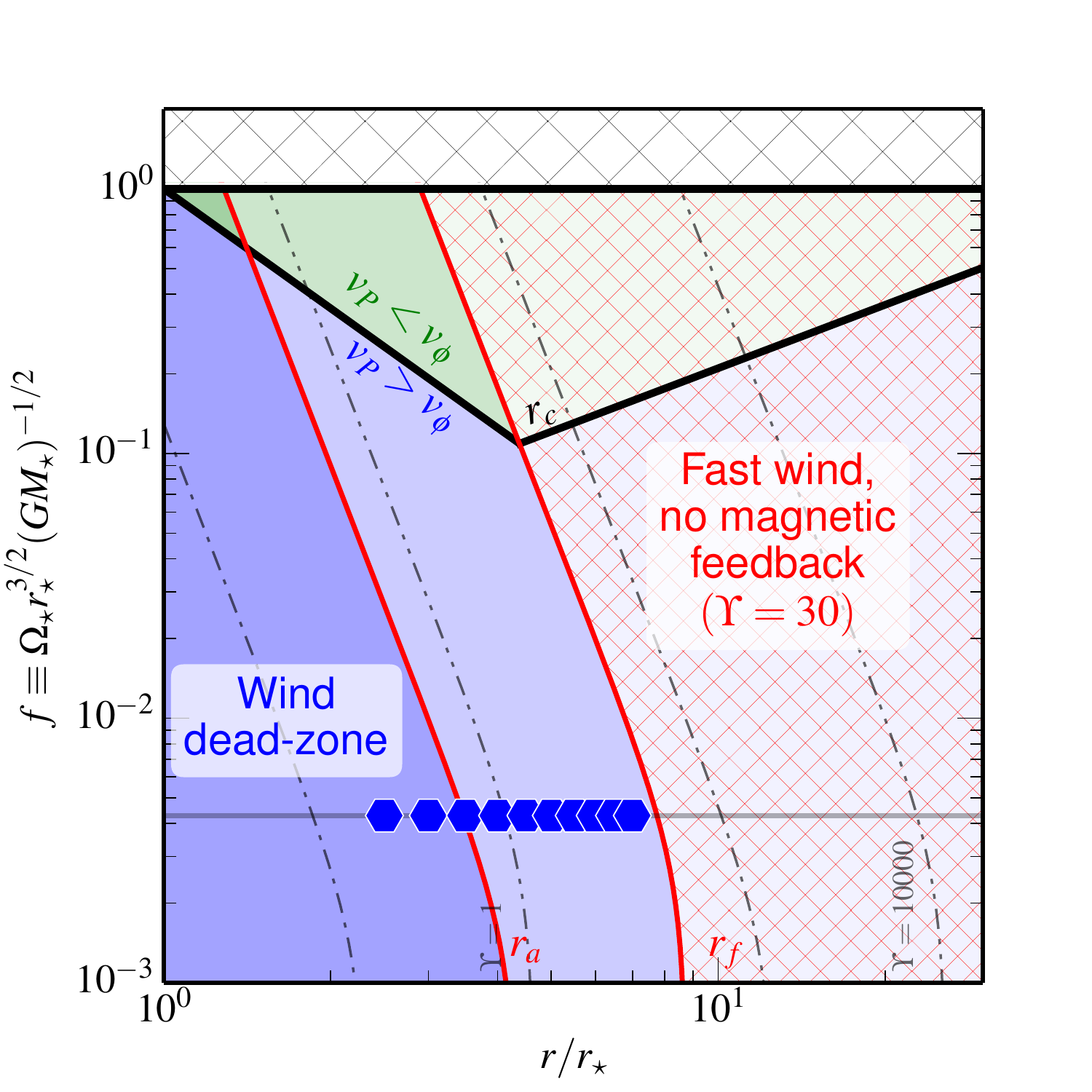}
  \caption{Wind characteristics as a function of rotation rate
    (vertical axis) and distance
  to the star (horizontal axis). The position of the Alfv\'en ($r_{a}$) and
  fast Alfv\'en ($r_{f}$) surfaces are shown in red for $\Upsilon=30$ and dashed-dotted
lines for $\Upsilon =1$ and $10^{4}$. The radius $r_{c}$ where the keplerian
velocity is equal to the \modif{azimuthal velocity of the
  wind} 
is labeled by the inclined black 
lines. The Alfv\'en surfaces delimit
the dead-zone and the fast wind regions. The top hatched region
corresponds to the unphysical case of
stars with surface velocities higher than the keplerian velocity. The
orbital radii considered in this paper are given by the blue hexagons
overlaid on the rotation rate of the fiducial star (horizontal grey line).}
  \label{fig:fig_schematic}
\end{figure}

The position of the Alfv\'en equatorial radii is illustrated in Figure
\ref{fig:fig_schematic}. The Alfv\'en surface position is shown as a
function of the fraction of break-up spin rate $f$ (vertical axis) and
magnetic confinement parameter $\Upsilon$ (pairs of oblique curves). The particular case of
$\Upsilon=30$ (red curves), which corresponds to the fiducial stellar wind
considered in this work, is
highlighted. 

The orbital velocity of the planet is
keplerian. For the sake
of simplicity, we assume a circular 
orbit for the planet. The orbital velocity only depends on the stellar
mass and the orbital distance and is given by $v_{P} =
\sqrt{GM_{\star}/r_{orb}}$ (under the approximation 
$M_{p}/M_{\star}\ll 1$). In Figure \ref{fig:fig_schematic} the black 
oblique lines represent the
radii $r_{c}$ at which the keplerian velocity is equal to the
\modif{rotational velocity of the wind.} 
It separates two interaction regions where the orbital
velocity of the planet is higher \modif{(blue areas) and lower (green
  areas)} than the azimuthal velocity
of the rotating wind. Any planet
orbiting outside the $r_{f}$ curve (red \modif{hatched} area) cannot magnetically influence the
star. A planet orbiting inside $r_{a}$ generally rotates much
faster than the wind itself (blue areas), although in the case of 
rapidly rotating stars the opposite situation may happen (green
areas). 

\reff{The SPMI is initially driven by the difference in azimuthal velocity
between the orbiting planet and the rotating wind. It is important to
realize that the orbital velocity of the planet $v_{P}$ is a function of
$r_{orb}^{-1/2}$, whereas the orbital angular momentum of the planet
is a function $r_{orb}^{1/2}$. If $v_P > v_\phi$, a positive transfer of angular
momentum from the planet to the star develops and leads to a decrease of
the orbital angular momentum. This decrease necessarily leads to an
increase of the planetary orbital velocity 
\citep[and a decrease of the
orbital radius, see aslo][]{Lovelace:2008bl,Laine:2011jt} as well as an increase in
the stellar rotation rate that should be accompanied by an increase
of the stellar wind rotation. }

\reff{Depending on the position of the planet
in the wind, two situations can occur. Inside the dead-zone, the
planet's orbital frequency usually increases faster than the stellar
rotation frequency (this is true for $r_{P}/r_{\star} \lesssim 0.5 (M_{\star}/M_{P})^{1/2}$)
As a consequence, the differential motion between
the planet and the wind increases and strengthens the SPMI. The 
planet migration associated with the SPMI is therefore
\textit{unstable}. A planet orbiting at the exact same rate as
the stellar wind should not transfer angular momentum magnetically to
the star, but the instability is such that any
perturbation would tend to grow and to make the planet
migrate. If the planet is inside the co-rotation radius (blue areas),
the SPMI leads to an orbital
decay of the planet until it reaches its Roche radius or simply merges
with its host. If the planet lies outside the co-rotation radius
(green areas), the
SPMI leads to an outward migration of the planet.}

\reff{Outside the
  Alfv\'en radius (red hatched zone), the azimuthal velocity of the
  wind falls off with cylindrical radius, and if the star rotates
  fast enough, a stable point exists where the planet is in
  co-rotation with the wind (black oblique line in the hatched
  zone). Of course, the stellar evolution, the tidal
interactions between the planet and its host, other rotating planets, or
even a disk should also be considered to determine the final
migration path of the planet \citep[see][for recent efforts to model
the evolution of such systems including the effects of tides]{Bolmont:2012go,Zhang:2014iz}.}

\reff{The positions of the Alfv\'en surfaces} are also plotted for much weaker and stronger magnetic
fields ($\Upsilon \in \{1, 10^{4}\}$) in
grey dash-dot and dash-double dot lines, for reference. The interaction
region where a planet
is able to provide magnetic feedback to its host grows with decreasing
stellar rotation rate and decreasing mass loss rate/increasing $B_{\star}$ (increasing
$\Upsilon$). This is a direct consequence of the shrinking/expansion of the
Alfv\'en surfaces in the equatorial plane (Equation \eqref{eq:ra_fit}). It must be stated here that
in reality, the magnetic
interaction regions are likely to be even more complex. We
have assumed here a dipolar topology for the stellar wind. For other magnetic
topologies the Alfv\'en surface on the stellar equator can be
pushed closer the stellar surface \citep{Reville:2014ud}. In the more
realistic case of
non-axisymmetric and/or cyclic magnetic fields, the radial location of the Alfv\'en
surface on the equator in the frame rotating with the orbiting planet
is time dependent \citep[see also][]{Pinto:2011ca}, which leads to a significant modulation of the
interaction regions.

Finally, the
magnetic interaction efficiency between a star and planet also strongly depends
on the planet characteristics, which will be explored in the following sections. 
In order to restrain the parameter space to study, we will focus 
here on one type of 
stellar wind labeled by the horizontal grey line in Figure
\ref{fig:fig_schematic} and vary the planet parameters (blue
hexagons). As can be seen in Figure \ref{fig:fig_schematic}, the results can be translated
accordingly for different stellar rotation rates and different mass loss rates.

\section{Planet models}
\label{sec:modeling-planets}

The axisymmetric geometry we consider is a
first step toward a more realistic modelling. It does not allow us
to \reff{fully describe} the intrinsically 3D star-planet
system\modif{: the planet is effectively represented by an axisymmetric torus
of large radius $r_{orb}$ and small radius $r_{P}$}. Nevertheless,
\citet{Cohen:2010jm} showed that even in the 3D global
geometry, a planet affects the stellar wind globally (\textit{i.e.}, at all longitudes).
In addition, our reduced 2.5D geometry captures
the basic physical ingredients of SPMIs, as will be made clear
in the two following Sections. \reff{As a first step, we chose the
axisymmetric 2.5D approach, in order to
explore a large parameter space, while still adequately modelling the
physical processes impacting the star.} These impacts are
likely to be overestimated in this study compared to a more realistic
3D case\modd{, which we leave for future work}.

\subsection{Simulation method and planet characteristics}
\label{sec:simulation-method-1}

We initialize a \modif{circular (in the $(\varpi,z)$ plane)} planet at
an orbital radius $r_{\rm orb}$ at the same time as the initialization
of the stellar wind. \modd{The planet's gravitational potential is added
  to the stellar potential in the whole domain.} The planet itself is \reff{modelled
as a boundary region in} which the pressure, density, velocity and
magnetic field have to be
prescribed. In all cases, we hold both the orbital radius and the
orbital velocity of the planet constant \citep[see][for more
details]{Strugarek:2014fr}. This is justified a posteriori by the fact
that the torques that develop in the system will make the planet
migrate over much longer time-scales than the overall simulated
temporal evolution. \modif{Given its proximity to its host star, the
  planet's rotation period is considered to be
  synchronized with the orbital period due to
  tidal locking.}
The system reaches a steady-state on a time scale of \modif{a few
  sound crossing times}. 
We checked that
doubling the resolution does not change our results by more than
3$\%$ in the magnetized-planet case (see Section
\ref{sec:dipolar-interaction}). \reff{In this case the planetary field
creates a shield around the planet that is barely affected by changes in the
resolution.} However, the non-magnetized planet cases (see
Section \ref{sec:unipolar-interaction}) show
a stronger dependency upon the grid resolution of our model: doubling
the overall resolution leads to variations of $30\%$ of the torque
applied by the planet to the star. \modif{This stronger resolution dependence is
expected} since, in the unipolar case, the resolution inside the planet determines the
\reff{level of numerical} dissipation in the planetary interior, which ultimately
counterbalances the induction, once a steady-state is reached. Even though the
numerical value of the torques can be affected by the grid resolution, the trends we
derive in Sections \ref{sec:magnetic-torques} and \ref{sec:discussion}
remain robust.

We choose typical
'hot-Jupiter' characteristics for our modeled 
planet. We set its radius to $r_P=0.1\, r_\star$ and its mass
to $M_P = 0.01 M_\star$. Considering spherical bodies and a circular
orbit for the planet, we can calculate the Roche limit
$d_{s,fl}=\alpha_{s,fl} r_{p}\left(M_{\star}/M_{P}\right)^{1/3}$, with
the coefficients $\alpha_{s}= 1.26$ in the solid case and
$\alpha_{fl}=2.44$ in the fluid case. The 
parameters we chose for our study lead to $d_{s}
= 0.59\, r_{\star}$ and $d_{fl} = 1.13\, r_{\star}$ in the solid and
fluid cases. Hence, all the orbital radii considered here lie well
outside the Roche limit of the star-planet system. Such a choice of parameters
corresponds \modif{closely} to the observed Corot-27 b \citep{Parviainen:2014hy}
or WASP-18 b \citep{Hellier:2009bk,Southworth:2009ca} planets.

 Following the terminology proposed by
\citet{Zarka:2007fo} \citep[see also][]{Kivelson:2004vf} in the context
of jovian satellites, we consider 
both the cases of \textit{unipolar} 
(weakly or non magnetized planet, hereafter labeled U) and \textit{dipolar} (strongly
magnetized planet with respect to the interplanetary medium, hereafter
labeled D) interactions. Those two cases could represent 
hot Jupiter
\citep{Lanza:2009fp} or Super-Earth \citep{Laine:2008dx} close-in
planets. We describe those two
types of interactions in detail in Sections
\ref{sec:unipolar-interaction} and \ref{sec:dipolar-interaction}
respectively. A list of all the simulations described in this work can
be found in Table \ref{tab:tab2}.

\begin{deluxetable*}{lccccccc}
  \tablecaption{Parameters and results of the SPMI cases\label{tab:tab2}}
  \tablecomments{$^{\dagger}$ Cases
    that oscillate between open and confined magnetic configurations.}
  \tablecolumns{8}
  \tabletypesize{\scriptsize}
  \tablehead{
    \colhead{Case} &
    \colhead{$r_{\rm orb}/r_\star$} &
    \colhead{$\theta_{0}$} & 
    \colhead{$B_{P}/B_w$ } &
    \colhead{$r_{\rm m}/r_P$ } &
    \colhead{$\tau^{\star}_{w}/\tau_{w}$} &
    \colhead{$\tau^{\star}_{P}/\tau_{w}$} &
    \colhead{$\tau^{P}/\tau_{w}$} 
  }
  \startdata 
U2.5 &  2.5 & \nodata & \nodata & \nodata &  0.85 & -3.50 &  2.96 \\
U3 &  3.0 & \nodata & \nodata & \nodata &  0.71 & -1.55 &  1.33 \\
U3.5 &  3.5 & \nodata & \nodata & \nodata &  0.70 & -0.95 &  0.86 \\
U4 &  4.0 & \nodata & \nodata & \nodata &  0.77 & -0.59 &  0.57 \\
U4.5 &  4.5 & \nodata & \nodata & \nodata &  0.83 & -0.26 &  0.30 \\
U5 &  5.0 & \nodata & \nodata & \nodata &  0.86 & -0.09 &  0.13 \\
U5.5 &  5.5 & \nodata & \nodata & \nodata &  0.94 & -0.01 &  0.04 \\
U6 &  6.0 & \nodata & \nodata & \nodata &  0.99 & $<$ 1\% & $<$ 1\% \\
U6.5 &  6.5 & \nodata & \nodata & \nodata &  1.00 & $<$ 1\% & $<$ 1\% \\
U7 &  7.0 & \nodata & \nodata & \nodata &  1.00 & $<$ 1\% & $<$ 1\% \\
\hline\\
D2a0 &  2.5 & $0$ &  7.5 10$^{0}$ &  1.7 &  0.89 & -1.14 &  1.36 \\
D2r0 &  2.5 & $\pi$ &  7.5 10$^{0}$ &  2.5 &  0.74 & -2.59 &  5.26 \\
D3a0 &  3.0 & $0$ &  3.4 10$^{1}$ &  1.4 &  0.69 & -0.40 &  0.40 \\
D3a1 &  3.0 & $0$ &  3.4 10$^{2}$ &  3.1 &  0.64 & -0.43 &  0.44 \\
D3a2 &  3.0 & $0$ &  8.4 10$^{2}$ &  4.6 &  0.56 & -0.46 &  0.48 \\
D3ia0 &  3.0 & $\pi/4$ &  3.4 10$^{1}$ &  1.6 &  0.84 & -0.02 &  0.04 \\
D3ia1$^\dagger$ &  3.0 & $\pi/4$ &  3.4 10$^{2}$ &  3.3 &  0.65 |  0.75 & -0.93 | -0.02 &  0.02 |  1.01 \\
D3ia2 &  3.0 & $\pi/4$ &  6.7 10$^{2}$ &  4.4 &  0.65 & -1.40 &  1.36 \\
D3i0$^\dagger$ &  3.0 & $\pi/2$ &  3.4 10$^{1}$ &  1.7 &  0.86 |  0.90 & -0.17 | -0.01 &  0.03 |  0.20 \\
D3i1$^\dagger$ &  3.0 & $\pi/2$ &  6.7 10$^{1}$ &  2.3 &  0.82 |  0.84 & -0.87 | -0.47 &  0.46 |  0.87 \\
D3i2 &  3.0 & $\pi/2$ &  3.4 10$^{2}$ &  3.5 &  0.67 & -2.32 &  4.04 \\
D3ir0 &  3.0 & $3\pi/4$ &  3.4 10$^{1}$ &  1.9 &  0.79 & -0.58 &  0.59 \\
D3ir1 &  3.0 & $3\pi/4$ &  1.0 10$^{2}$ &  2.7 &  0.83 & -1.86 &  2.57 \\
D3ir2 &  3.0 & $3\pi/4$ &  3.4 10$^{2}$ &  4.5 &  0.77 & -2.99 &  4.60 \\
D3r0 &  3.0 & $\pi$ &  3.4 10$^{1}$ &  1.8 &  0.73 & -0.46 &  0.84 \\
D3r1 &  3.0 & $\pi$ &  6.7 10$^{1}$ &  3.3 &  0.75 & -1.23 &  2.33 \\
D3r2 &  3.0 & $\pi$ &  3.4 10$^{2}$ &  6.2 &  0.77 & -3.22 &  5.83 \\
D4.5a0 &  4.5 & $0$ &  1.7 10$^{3}$ &  3.8 &  0.87 & -0.11 &  0.32 \\
D4.5r0 &  4.5 & $\pi$ &  1.0 10$^{3}$ &  5.0 &  0.72 & -0.32 &  1.06 \\
D6a0 &  6.0 & $0$ &  2.0 10$^{3}$ &  4.8 &  0.99 & $<$ 1\% &  0.21 \\
D6a1 &  6.0 & $\pi$ &  6.1 10$^{2}$ &  5.2 &  0.93 & -0.01 &  0.33 
  \enddata
\end{deluxetable*}

\subsection{Unipolar interaction}
\label{sec:unipolar-interaction}

The unipolar interaction refers to an interaction of a magnetized
medium with a weakly (or non-) magnetized obstacle. In the context of
SPMI, the magnetized medium is the stellar wind and the obstacle is
the planet. This interaction was initially modeled in the context of the
Io-Jupiter system by \citet{Goldreich:1969kf}. Several unipolar interaction
cases need to be distinguished. 

In the solar system, Venus provides an example of unipolar
interaction between a planet and the wind of its host star. Although
Venus possess no intrinsic magnetic field, it has
a neutral atmosphere that is efficiently screened from the surrounding
solar wind by its ionosphere. More precisely, the very high ionospheric conductivity
prevents the solar wind magnetic field from permeating into the atmosphere of
Venus \citep{Russell:1993jk}. Such unipolar cases lead to the
creation of an induced magnetosphere in the planet vicinity, which
possess the same global \modif{structure --from the point of view of
the wind--} as the self-generated
\citep[via an internal dynamo, see][]{Stevenson:2003ji}
magnetospheres of planets like the Earth or Jupiter, although the
induced magnetospheres owe
their origin to a completely different process \modif{and are generally much
less spatially-extended} \reff{\citep[for a recent modelling of the
magnetosphere of Venus and its interaction with the solar wind
plasma, see][]{Ma:2013aa}}.

It must be noted, however, that a planetary
ionosphere does not always provide an effective magnetic shield to the ambient
stellar wind. The ionospheric barrier can indeed break
if the stellar wind is sufficiently dense, sufficiently fast, or if
the stellar \modif{ionizing influence (through either ion pick-up \reff{or}
  high energy radiation)} is sufficiently low \citep{Russell:1993jk}. Such conditions can be
realized for close-in exoplanets that interact with a much more dense
stellar wind than distant planets do. In this case, the wind magnetic
field can permeate 
into the planetary interior. \reff{The interaction that develops in this
case is the so-called generalized Alfv\'en wings scenario
\citep{Neubauer:1998aa}. We consider here a simplified, ideal MHD
formulation of the problem, in which the Alfv\'enic perturbations
associated with the planet are either reflected and absorbed at the
stellar boundary, or travel to the outer boundaries of the
domain. Models taking into account a finite conductivity of the
ambient plasma, where the waves can be reflected in between the two
bodies, can be found in, \textit{e.g.},
\citet{Neubauer:1980in,Kivelson:2008aa}. In addition, because of the
axisymmetric geometry we consider here, any reflected perturbation will
always come back to the orbiting planet and the planet will behave as
the so-called unipolar inductor \citep{Goldreich:1969kf}.}
Then, two extreme cases may occur
\citep{Laine:2008dx}. If the planetary electric conductivity is very
high compared to the stellar surface conductivity, magnetic
field lines are frozen in the planet and dragged as the
planet orbits \citep{Laine:2011jt}. Conversely, if it is very low,
the wind magnetic field
diffuses through the planet. For moderate conductivities,
one would expect that both effects play a role in the SPMI. A slipping time 
can be defined by the time it takes a magnetic
flux tube to slip through the planet. The slipping time depends on
the relative orbital motion of the planet in the ambient rotating
wind, and on the ratio of the electric conductivities between the planetary
interior and the stellar surface. The reduced 2.5D axisymmetric geometry we chose for
this first study allows us to model only the case where the planet drags the magnetic field
as it orbits \reff{\citep[similar to the Io-Jupiter case, with a closed
  current circuit between the planet and its host, see][]{Strugarek:2012th}}. 

A full treatment of the various unipolar cases would
require the description of the ionisation of the planetary
atmosphere by the stellar wind and stellar UV radiation, which is beyond the scope of this
study. This effect, combined with diffusive effects acting in the
planet interior, is left for
future work. Furthermore, the geometry in the induced magnetosphere case cannot be
realistically modeled with a 2.5D axisymmetric setup; hence, we will
focus our study on the unipolar case with no \modif{induced} magnetosphere.

\begin{figure*}[tbp]

  \includegraphics[width=0.44\linewidth]{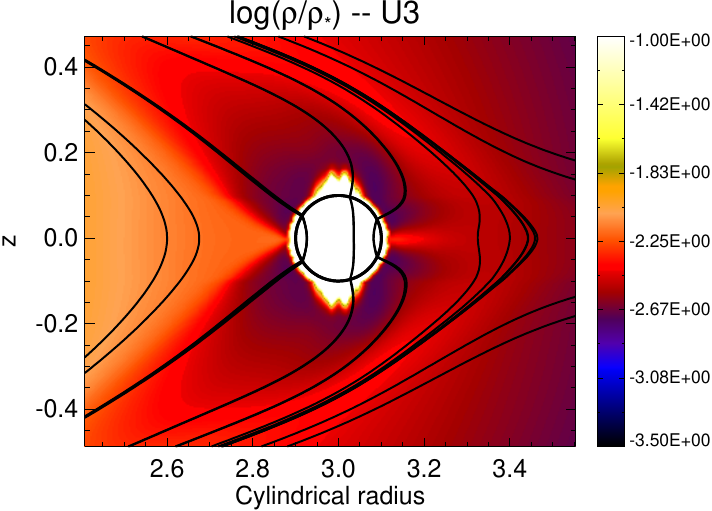}
  \includegraphics[width=0.56\linewidth]{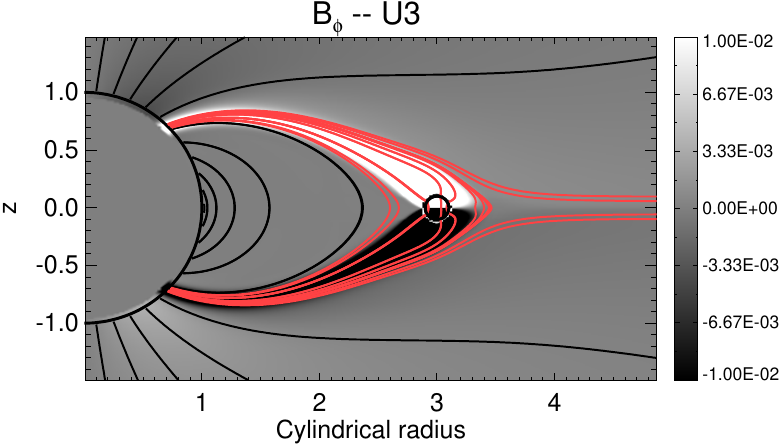}

  \caption{Unipolar interaction case U3. 
    The planet is located at
    $r_{\rm orb} = 3\, r_\star$. The left panel represents the plasma
    density (in log scale) \modif{normalized to the stellar coronal
      base density}. The planet surface is shown by a black
    circle. The solid black lines
    are magnetic field lines.
    The right panel shows the azimuthal
    component of the magnetic field (black and white map). The solid lines are the poloidal
  magnetic field lines, which are
  emphasized in red in the planet vicinity.}
  \label{fig:fig_unipolar_example}
\end{figure*}

We let the magnetic field freely evolve inside the planet and 
fix all the other variables. The density profile inside the planet
then determines the development of the SPMI. We tested various radial
shapes of the density profile (from constant to gaussian) under the
constraint of a given planetary mass. We found that the density
jump between the planet surface and the ambient wind is the important
control parameter for the level of interaction the system is able to
reach. The detailed shape of the density profile inside the planet only
marginally affects the SPMI properties. We varied the density
contrast between the planet and the ambient stellar wind 
from $10^{5}$ to $10^{20}$. \modif{The torques that develop in the
  unipolar cases (results shown
in Section \ref{sec:magnetic-torques}, see Figure
\ref{fig:torq_magscan}) depend} 
linearly on the logarithm of this density contrast. On one
hand, the modification of the stellar wind \modif{torque} increases only by $3\%$
between density ratios from $10^{5}$ to
$10^{20}$ and can hence be neglected. On the other hand, the torque
applied by the planet to the star increases (in absolute value) by $25\%$
due to the amplification of the azimuthal magnetic field in the flux
tube \modif{linking the star and the planet}. We choose in the
remainder of this paper to consider a density
contrast of $10^{13}$. Our results are robust to small (factor of $~2$) variations of the density contrast, but larger variations have
to be taken into account when using the torque scalings that will be derived in
Section \ref{sec:magnetic-torques}.

Figure \ref{fig:fig_unipolar_example} displays
the solution we obtained for case U3 (see Table \ref{tab:tab2}), with a
planet located at $3\, r_\star$ (just inside
the dead-zone of the stellar wind). We show the density (left panel)
in the a steady-state solution. A very thin, higher-density boundary
layer is created around the planet. It does not pollute the wind plasma
and its size is robust with respect to both the grid resolution and
the density contrast. On the right panel the azimuthal magnetic field
is shown by the black and white color map. The magnetic field lines
connecting the planet to the star (and nearby field lines) are highlighted in red. We observe
that the magnetic flux-tube is strongly elongated in the azimuthal
direction, due to fast orbital motion of the planet in the relatively
slowly rotating
wind. \modif{The pitch-angle $|B_{\phi}/B_{p}|$ in the flux tube
  reaches values of $\sim 3$ on the dayside of the planet. On the
  nightside, it reaches occasionally larger values ($<10$) that lead to
  a fast outward opening and reconnection of the magnetic field lines. The
  small magnetic lobes observed on the left panel in each hemisphere on the
  nightside of the planet are residuals from those occasional
  reconnection events. These events are reminiscent from the
  well known twisting/slipping mechanism \citep[see,
  \textit{e.g.},][in the context of star-disk
  interaction]{Matt:2005ab} and are a trace of the finite amount of dissipation
  imposed by our grid. The
  time dependence introduced by these events in our simulations are
  observed to have a negligible influence on the global properties of
  the star-planet system.}

The boundary condition on pressure then determines whether or
not some plasma escapes away from the surface of the planet. Because we are
interested primarily in angular momentum transfers between the star
and the planet, we
chose here to neglect atmospheric escape and design the planet boundary
conditions such that no outflow from the planet is generated. The same
care was taken for the dipolar interaction cases, which we describe
in the next section.

\subsection{Dipolar interaction}
\label{sec:dipolar-interaction}

We enforce
a dipolar field at the planetary surface in dipolar cases. We adapted
the magnetic dipole 
formula for our planet --which is shifted from the origin of the study
frame-- so that it still satisfies 
$\boldsymbol\nabla\cdot\mathbf{B}=0$. Such a dipolar field is given by
\begin{eqnarray}
 \label{eq:dipole_cyl_pluto_br_planet}
 B_{\varpi} &=& \mu_{P}\cos{\left (\theta_0 \right
  )}\frac{3z\varpi'}{\left(\varpi'^{2} + z^{2}\right)^{\frac{5}{2}}}
\nonumber \\ &+& \mu_{P}\sin{\left (\theta_0 \right
  )}\frac{2z^{2}-\varpi'^{2}}{\left(\varpi'^{2}
    +z^{2}\right)^{\frac{5}{2}}}\, , \\
 \label{eq:dipole_cyl_pluto_bz_planet}
 B_{z} &=& \mu_{P}\cos{\left (\theta_0 \right
  )} \frac{z^{2}\left(2-\frac{r_{\rm
        orb}}{\varpi}\right)-\varpi'^{2}\left(1+\frac{r_{\rm orb}}{\varpi}\right) 
}{\left(\varpi'^{2} +z^{2}\right)^{\frac{5}{2}}} \nonumber\\  &+& \mu_{P}\sin{\left (\theta_0 \right )}\frac{ -3\varpi' 
        z + \frac{z}{\varpi} \left(\varpi'^{2} +
          z^{2}\right)}{\left(\varpi'^{2}+z^{2}\right)^{\frac{5}{2}}}
\end{eqnarray}

where $\varpi'=\varpi-r_{\rm orb}$, $\mu_P$ is the dipolar moment
of the planetary magnetic field and $\theta_{0}$ is the tilt angle of the
planetary dipole with respect to the vertical axis. The
interaction of the magnetosphere with the coronal wind establishes a
steady-state planetary magnetosphere of finite size. In all cases, the
velocity of the planet is set to be keplerian, and the other velocity components
are set to zero at the surface of the planet. We define each case by
the ratio of the planetary field at its pole
($B_{P}=\mu_{P}/r_{P}^{3}$) to the local wind magnetic field $B_{w}$
(see table \ref{tab:tab2}).

\begin{figure*}[tbp]

  \includegraphics[width=0.44\linewidth]{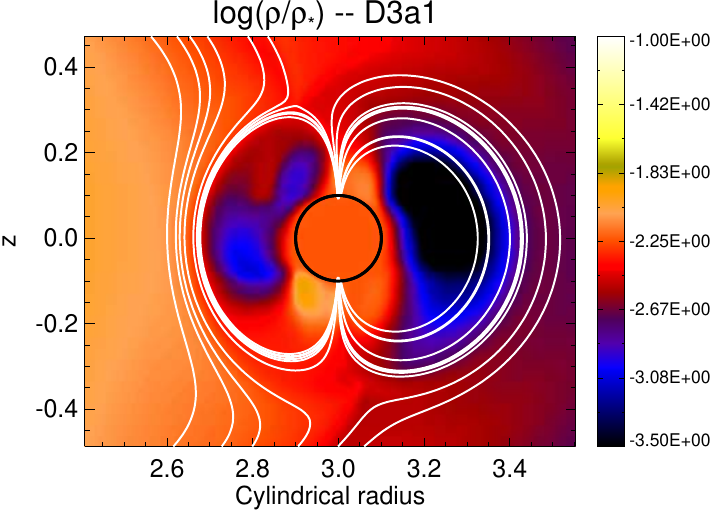}
  \includegraphics[width=0.56\linewidth]{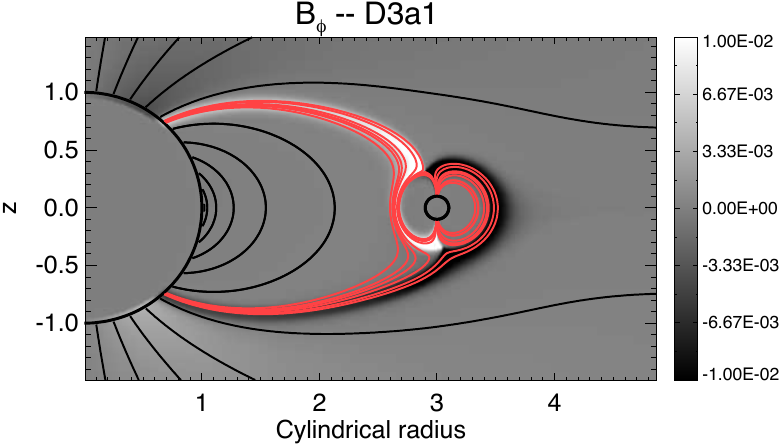}
  \includegraphics[width=0.44\linewidth]{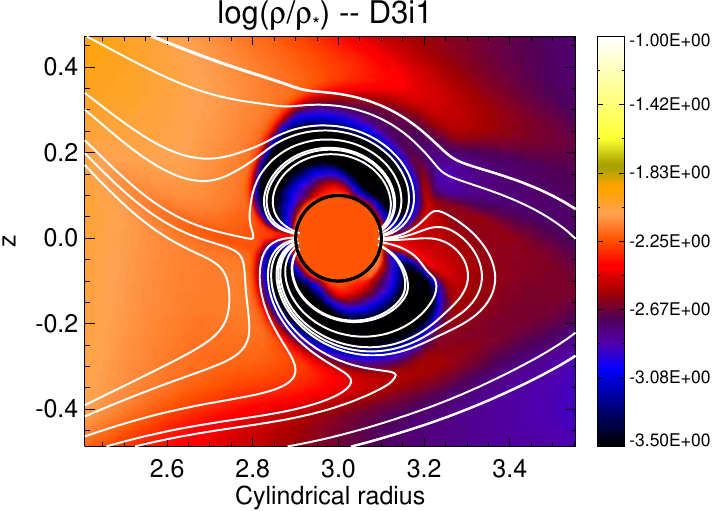}
  \includegraphics[width=0.56\linewidth]{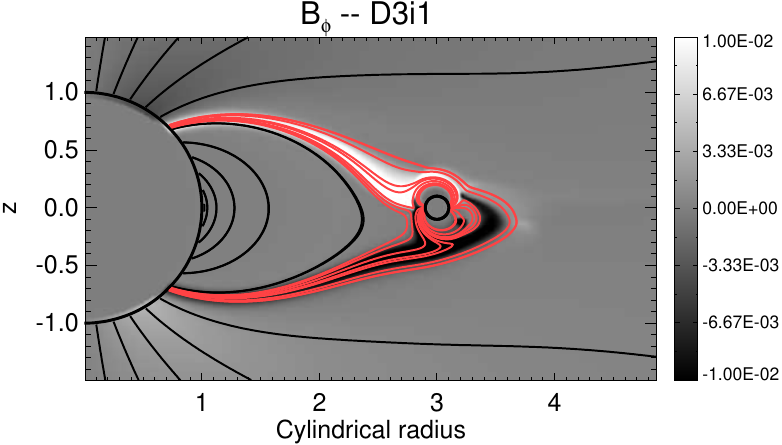}
  \includegraphics[width=0.44\linewidth]{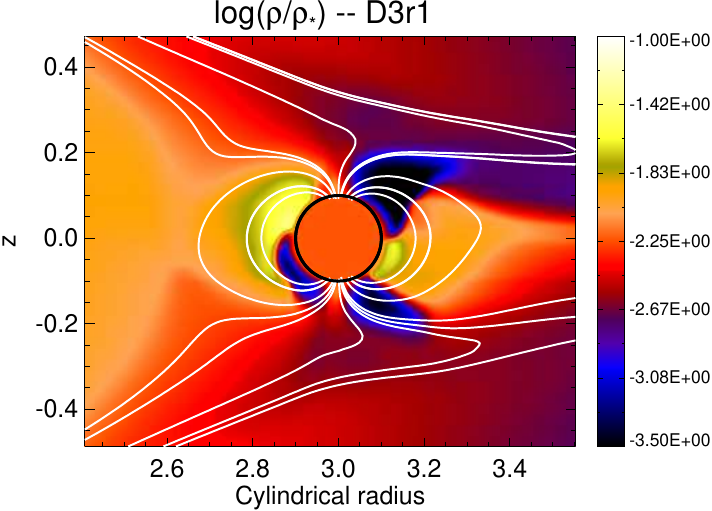}
  \includegraphics[width=0.56\linewidth]{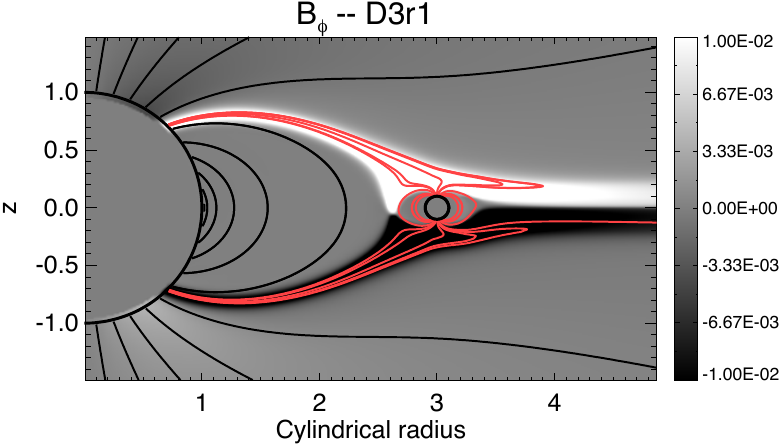}
  \caption{Dipolar interaction solutions for $\theta_{0}=0,\, \pi/2$ and
    $\pi$ with $r_{\rm
      orb}=3\, r_{\star}$ \modif{(from top to bottom). The quantities
    shown and layout are the same as in Figure \ref{fig:fig_unipolar_example}.}}
  \label{fig:fig_dipolar_example}
\end{figure*}

In Figure \ref{fig:fig_dipolar_example} we display three 
dipolar interaction cases, for a planet located at $r=3\, r_\star$ (inside
the dead-zone of the stellar wind), with
different magnetospheric angles $\theta_{0}$ (see Table \ref{tab:tab2} for the
cases parameters). We observe that, in all three cases,
the planet is able to retain a magnetosphere, whose
finite size depends on $B_{P}/B_{w}$, $\theta_{0}$, and on the position of
the planet in the wind. \modif{We define the size of the planetary
  magnetosphere $r_{m}$ by the extend of the last closed magnetic
  field line of the planet on the planetary magnetic
  equator, on the dayside. The magnetospheric sizes for all the dipolar cases are
  shown in table \ref{tab:tab2}.}

In all cases, there is a direct magnetic link
from the magnetic poles of the planet to the star. For
the wind considered here and a planet located at $r_{orb}=3\,
r_{\star}$, \modif{the foot-point of the magnetic link on the stellar surface is located near
the open-closed field lines transition region.}
The SPMI develops the same qualitative behavior in the unipolar and
dipolar cases: elongated field lines with a strong positive
(resp. negative) azimuthal magnetic field in the north (resp. south)
hemisphere connect the star to the planet. The qualitative resemblance
of all the SPMI cases suggests
that the shape of the magnetic interaction between the two bodies is
primarily determined by the strong coronal stellar magnetic field.

We observe nonetheless significant differences in the three dipolar
cases displayed in Figure \ref{fig:fig_dipolar_example}. Some magnetic
field lines originating from
the poles of the planet open in the stellar wind in the anti-aligned ($\theta_{0}=\pi$)
case.
It is now well established that the direction of
the planetary magnetic field naturally leads to a 'closed' or an 'open'
interaction case \citep{Ip:2004ba}. The closed configuration (upper
panels in Figure \ref{fig:fig_dipolar_example}) efficiently confines
most of the plasma inside the \modif{planetary} magnetosphere, and leads to a very thin
magnetic link between the two bodies. The resulting magnetic
configuration of the planet is stable with respect to external perturbations by
the stellar wind. Conversely, the open case (lower
panels) leads to a much wider magnetic link in the polar region of the
planet magnetosphere. The magnetic link in this case is 
sensitive to external perturbations by the wind. 
By considering
$r_{orb}=3\,r_{\star}$, the magnetosphere of
the planet is sufficiently close to boundary of the streamer
(the open-closed field lines transition region in the stellar wind) 
that the wind is able to perturb some of the magnetic
field lines connecting the two bodies. \modif{In the aligned case the
  magnetic configuration is strong enough to resist this wind
  perturbation and remain in the closed configuration. On the contrary,
  the wind is able to drag some of the connecting magnetic field lines
  in the anti-aligned case, which leads to the observed open
  configuration.} The resulting \modif{radially} elongated field
lines are then forced to \modif{episodically reconnect and re-open} on the nightside
\modif{of the planet}, as seen in the
lower right panel of Figure \ref{fig:fig_dipolar_example}. This
phenomenon will provide an additional source of angular momentum loss
for the planet. 
The inclined magnetosphere case (middle panels) lies in
between those two configurations. In the cases shown, the planetary field is
sufficiently small that its magnetosphere is
confined inside the dead-zone \modif{and remains in a closed
  configuration}. For more vigorous fields (case D3i2,
see Table \ref{tab:tab2}), the inclined magnetosphere also opens into
the wind akin to the reversed configuration. Finally, some of the inclined
cases regularly flip from a confined to an open
configuration. Provided the planetary field is sufficiently small or
large, though, one of the two steady configurations is systematically obtained.

The plasma density in the magnetosphere also differs
from one case to the other. The magnetospheric plasma has low density
in the aligned and inclined cases, with plasma
concentrations in the polar
regions. The \modif{anti-aligned} case shows a more complex density structure in the
magnetosphere with density concentrations in the magnetic
equatorial regions. The depleted regions coincide with the open/closed field
line interfaces of the planetary magnetosphere. The detailed density structure
here is likely to depend upon the details of the reconnection process
in the nightside \reff{that is likely affected by the 2.5 reduced
  by the grid resolution}.
 We observed, by refining the grid,
that the overall properties of the star-planet system (such as
magnetic torques and angular momentum transfers, see Section
\ref{sec:magnetic-torques}) were only marginally influenced by the
detailed density structure of the planetary magnetosphere. A
more accurate modelling of the planetary magnetosphere configuration would
require a better control of the reconnection process in the nightside
of the magnetosphere.

\section{Magnetic torques}
\label{sec:magnetic-torques}

We now quantify the effects of SPMIs \reff{for a range of planetary}
orbits inside the Alfv\'en surface, and for both 
unipolar and dipolar interactions. 
\modif{The Lorentz force associated with the magnetic link between the
  star and the planet leads to a magnetic angular momentum transfer,
  as well as a modification of the stellar wind, which we describe in 
Section \ref{sec:angul-moment-transf}}. We \modif{characterize}
the effects of the planetary
magnetic field amplitude and inclination (Section
\ref{sec:incl-plan-magn}), and the effects of the
position of the planet in the wind (Section
\ref{sec:unip-dipol-inter}) on the torques associated to the SPMI. 
Finally, we estimate the
planet migration induced by the SPMI in Section \ref{sec:planet-migration}. 

\subsection{Angular momentum transfers and stellar wind modification}
\label{sec:angul-moment-transf}

\begin{figure}[tbp]
  \includegraphics[width=\linewidth]{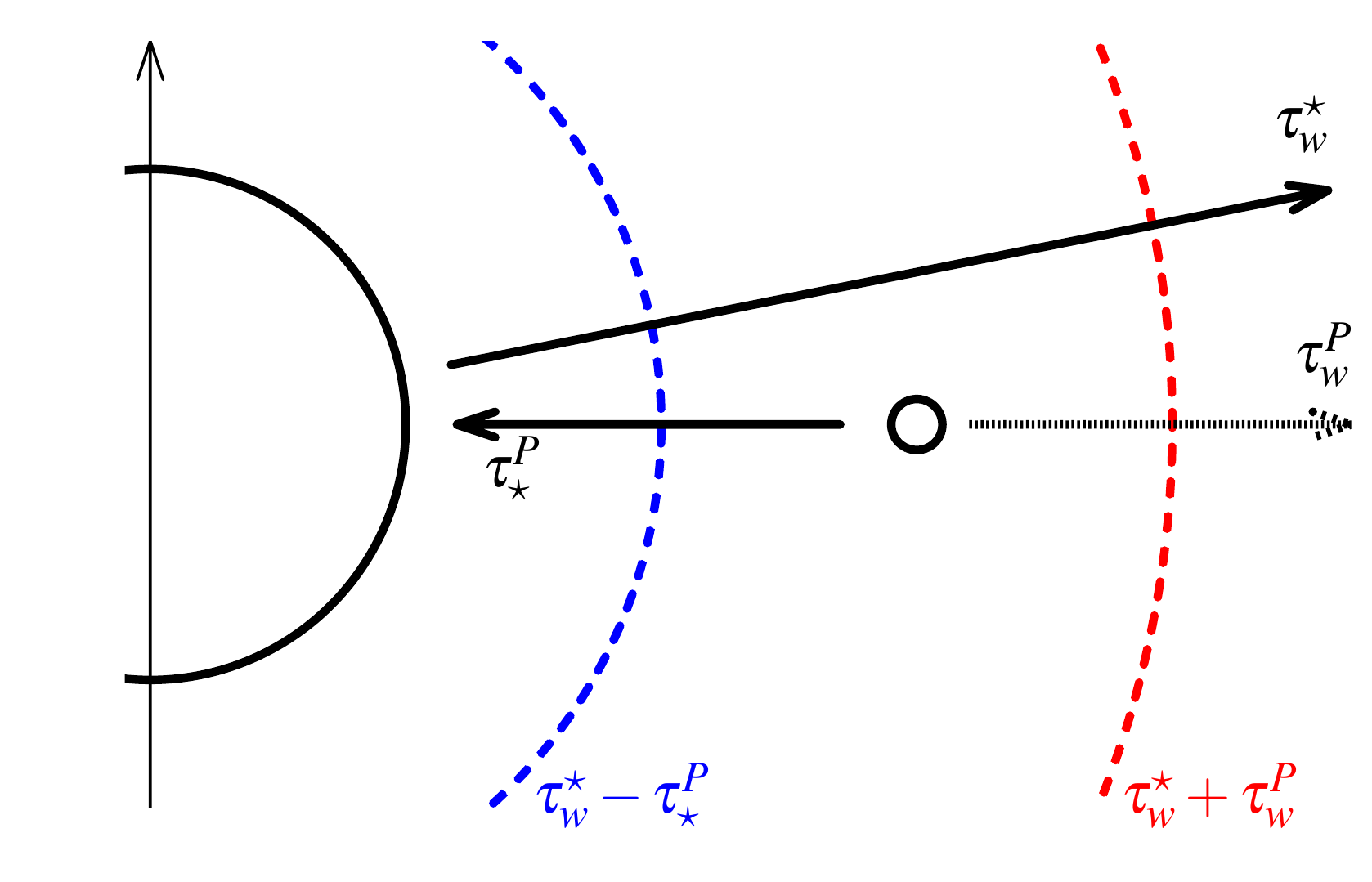}
  \includegraphics[width=\linewidth]{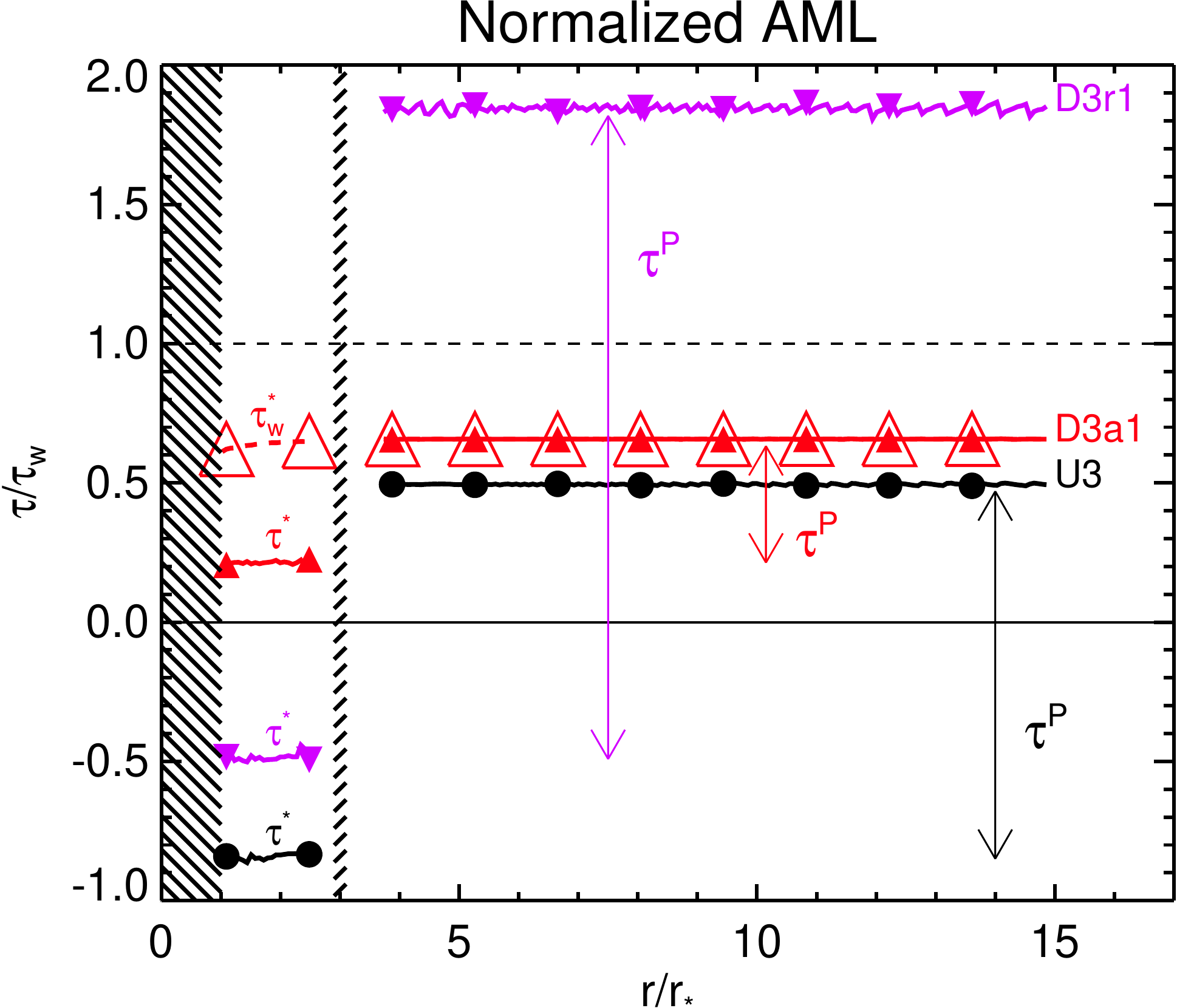}
  \caption{\textit{Top.} Schematic of the angular momentum transfers in
    a star-planet system. \reff{The black arrows show the direction of the
    angular momentum flux in the cases studied here. They are labeled
    with the physical sources from which they originate.} \textit{Bottom.}
  Angular momentum loss rate
    (Equation \ref{eq:angmom_flux} \modif{integrated
      over spheres})
    normalized to the fiducial wind AML $\tau_{w}$, as a function of the spherical radius.
    The star ($0\le r \le r_{*}$) and planet ($r_{\rm orb}=3\, r_{*}$) positions are labeled by the
    hatched zones. \reff{Three cases are shown: a unipolar case (U3, black
    lines), a closed dipolar case (D3a1, red lines) and an open
    dipolar case (D3r1, magenta lines). The torque applied to the
  star $\tau^{\star}$ and applied to the planet $\tau^{P}$ are indicated. The
  dashed red line shows the outward AML only in case
  D3a1.}}
  \label{fig:fig_examples_amom_r}
\end{figure}

The magnetic connection between the star and the planet leads to a magnetized
angular momentum transfer. Since we fix the stellar
rotation rate and the orbital motion, the planet and the star
act as source and sinks of angular momentum, which is conserved in the
stellar wind \citep{Strugarek:2014fr} and everywhere on our
computational grid. As a result, when the system has reached a
steady-state, the flux of
angular momentum integrated on any spherical suface that is not crossing the
stellar or the planetary interior is constant.  The top
panel of Figure \ref{fig:fig_examples_amom_r} is a schematic of
magnetized angular
momentum transfers in a star-planet system. The angular momentum
flux in between the star and the planet (crossing the dashed blue line)
includes both the torque
applied by the planet and the torque applied by the wind to the
star. It thus corresponds to the overall angular momentum extracted from
the star, which we denote
\reff{$\tau^{\star}=\tau^{\star}_{P}+\tau^{\star}_{w}$, with
  $\tau^{\star}_{P}$ the torque applied by the planet to the star and
  $\tau^{\star}_{w}$ the torque applied by the wind to the star. If the star
is rotating slowly enough (as it is the
case here), the angular momentum is always transfered from the
  planet to the star and is associated with the torque applied to the
planet by the star, which we denote
$\tau^{P}_{\star}=-\tau^{\star}_{P}$. 
As a consequence, when
we compute the AML between the star and the
planet, we can account separately
for positive and negative contributions and thus properly separate
the stellar AML associated with the wind
($\tau^{\star}_{w}$) from the one
 associated with the planet ($\tau^{\star}_{P}$).} The angular momentum flux
outside the orbital radius (crossing the dashed red line) results
from the torque applied
by the wind to the star \reff{($\tau^{\star}_{w}$) and to the planet
  ($\tau^{P}_{w}$)}. By subtracting the inner
and outer AMLs, one readily obtains the 
\reff{torque applied to the planet
$\tau^{P} = \tau^{P}_{\star} + \tau^{P}_{w}$}. The wind contribution
to the planetary AML is labeled with dots
because in some SPMI cases, the planet does not lose any angular
momentum to the wind.

\reff{The bottom panel of Figure \ref{fig:fig_examples_amom_r} shows the time-averaged AML
(Equation \ref{eq:angmom_flux} normalized to the fiducial wind AML
$\tau_{w}$) 
as a function of the spherical radius, for three typical cases (U3,
D3a1 and D3r1, see Table \ref{tab:tab2}).} As expected,
the AML is a piecewise constant function of $r$, which confirms 
\textit{(i)} the conservation of angular momentum and
\textit{(ii)} that the models have reached a statistical steady-state. 

We immediately remark that the total torque applied to the star
\modif{(curves lying in $r<3\,r_{\star}$)} is
strongly reduced compared to the fiducial wind torque (black thin line).
In some cases, the sign of the net torque on the star is 
even reversed: the connection between the star
and the planet is strong enough such that the net torque is
accelerating the star. We recall here that our 2.5D
axisymmetric setup overestimates 
these torques (see discussion in Section \ref{sec:unip-dipol-inter}),
so the quantitative values of the torques we
obtain here need to be checked with 3D simulations. 

\begin{figure}[tbp]
  \subfigure[]{
    \includegraphics[width=\linewidth]{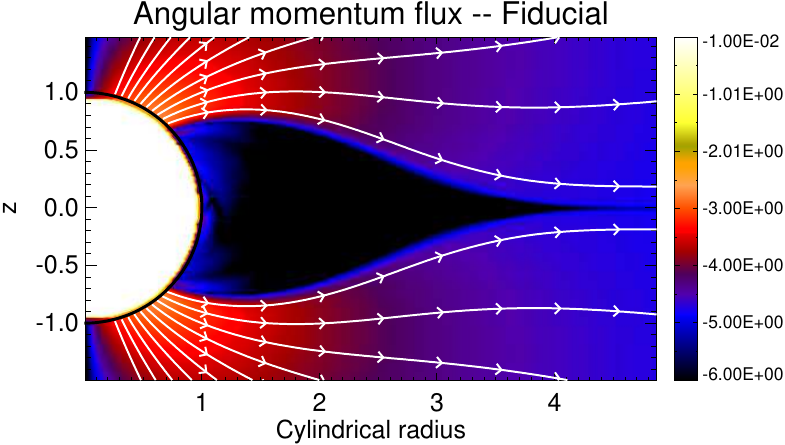}
  }
  \subfigure[]{
    \includegraphics[width=\linewidth]{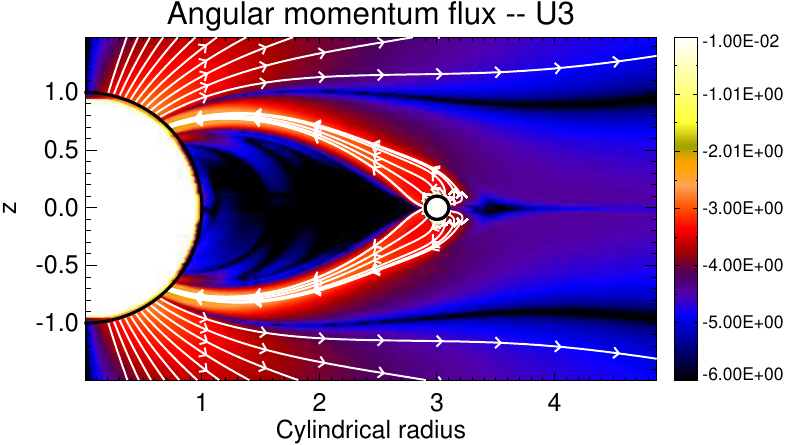}
  }
  \subfigure[]{
    \includegraphics[width=\linewidth]{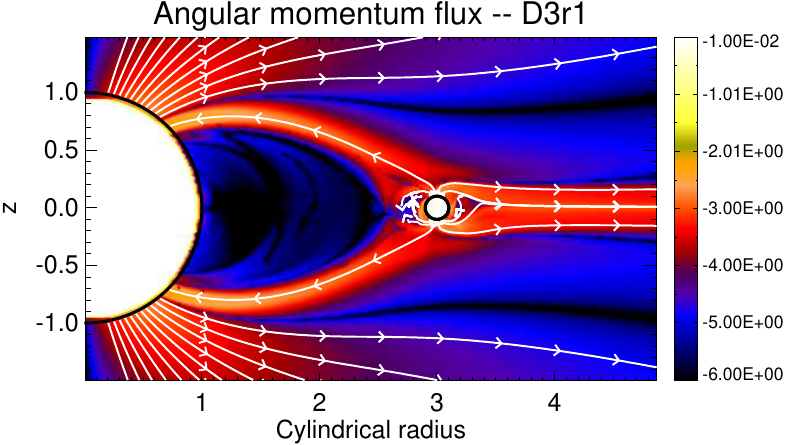}
  }
  \caption{Two dimensional angular momentum fluxes. The streamlines of
    angular momentum are labeled by the white lines and arrows, its
    amplitude is shown by the background colormap (in logarithmic
    scale). The planet and stellar surfaces are indicated by the black
    circles. The cases of (a) the fiducial stellar wind, (b) unipolar and
    (c) dipolar ($\theta_{0}=\pi$) interactions are shown.}
  \label{fig:fig_example_amomflux}
\end{figure}

The red dashed line (with open symbols) shows the $\tau^{\star}_{w}$
for the aligned case D3a1. We see that the filtered AML is
\modif{indistinguishable from}
the AML calculated for $r>r_{orb}$ which means that --in this
particular case-- the only torque applied to the planet comes from its
magnetic connection to the star. Conversely, we observe that the
torque applied to the star-planet system is larger than the fiducial
torque in case D3r1, which indicates that the planet is losing angular
momentum to both the star and the wind. In order to visualize those direct
magnetized transfer of angular momentum, we display the 2.5D
angular momentum fluxes in Figure \ref{fig:fig_example_amomflux} for the
fiducial wind and for cases U3 and D3r1.
The direction of the angular momentum flux is labeled by the white stream lines and its
amplitude by the logarithmic colormap. The angular momentum naturally
flows out of the star in the open field lines regions (panel a) when
no planet orbits around the star. In panels (b) and (c), angular
momentum directly flows from the planet to the star as well, following the
azimuthally elongated 'flux-tube' created by the SPMI
(see Figures \ref{fig:fig_unipolar_example} and
\ref{fig:fig_dipolar_example}). The wind driving is
modified at its foot-point on the stellar surface. The \modif{size of
  the} open field
line region is accordingly diminished compared to the case of a planet-free
wind (panel a) --which explains the general decrease of the wind torque.

The 2.5D representation of the angular momentum fluxes
illustrates the importance of magnetic topology in SPMIs. The unipolar case does
not add any fundamental constraint on the stellar wind magnetic
topology: the magnetic field lines are simply dragged in the
azimuthal direction by the orbiting planet. 
The anti-aligned case displayed in panel (c) develops the same
kind of connection to the star, but the planetary magnetic field is
also able to open up in the accelerating wind region. This interaction
is made possible by the closeness of the planet to the dead-zone
boundary on the equator and the anti-aligned topology of the
planetary magnetic field (see Section
\ref{sec:dipolar-interaction}). Hence --as expected from
Figure \ref{fig:fig_examples_amom_r}-- the planet loses angular momentum
to both the wind and the star at the same time. We now explore the
quantitative sensitivity of the SPMI to the planetary magnetic field strength and
topology (Section \ref{sec:incl-plan-magn}), and to the planet
position inside the Alfv\'en surface (Section \ref{sec:unip-dipol-inter}).

\subsection{Topology and strength of the planetary magnetic field}
\label{sec:incl-plan-magn}

\begin{figure}[tbp]
    \includegraphics[width=\linewidth]{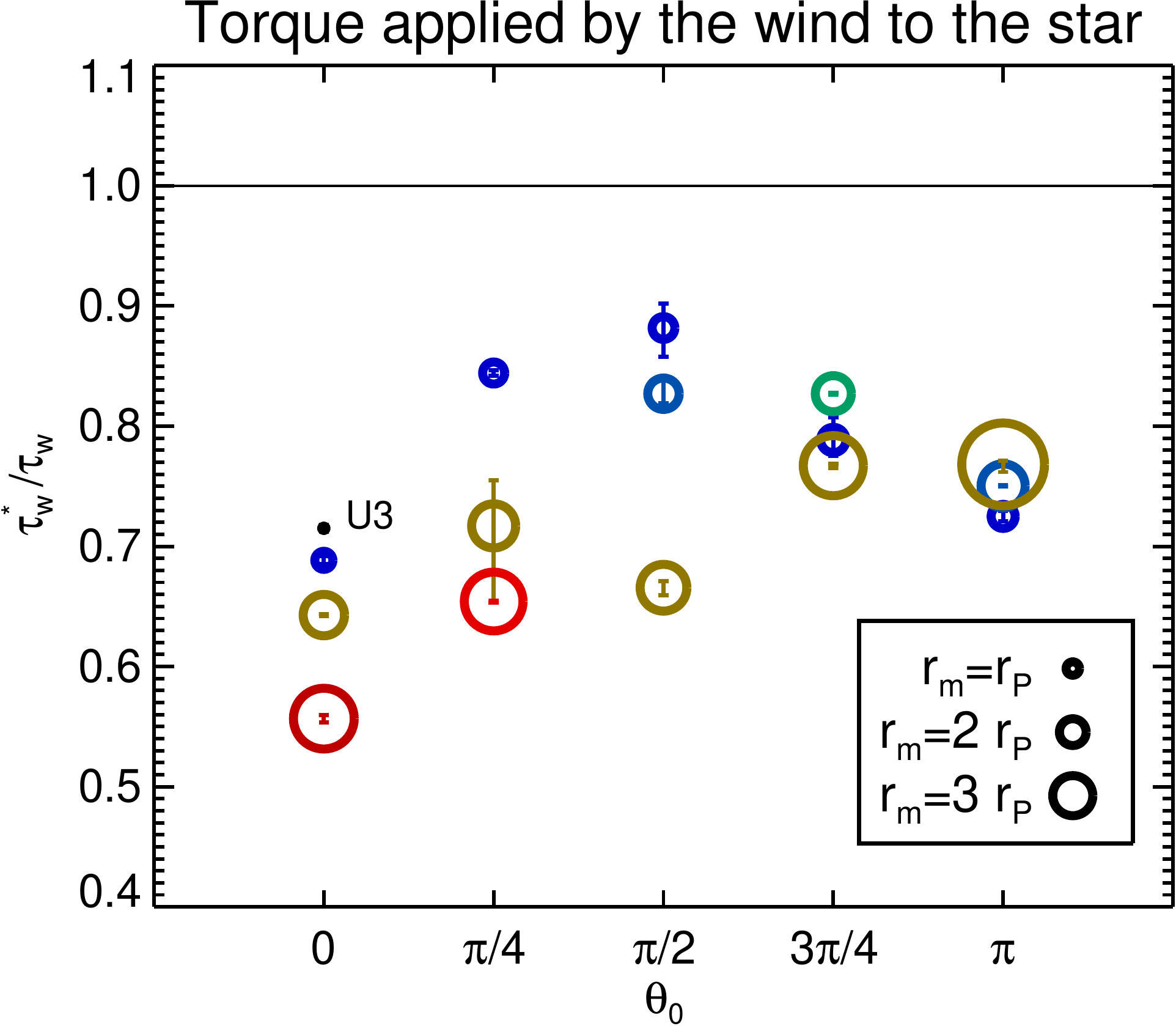}
    \includegraphics[width=\linewidth]{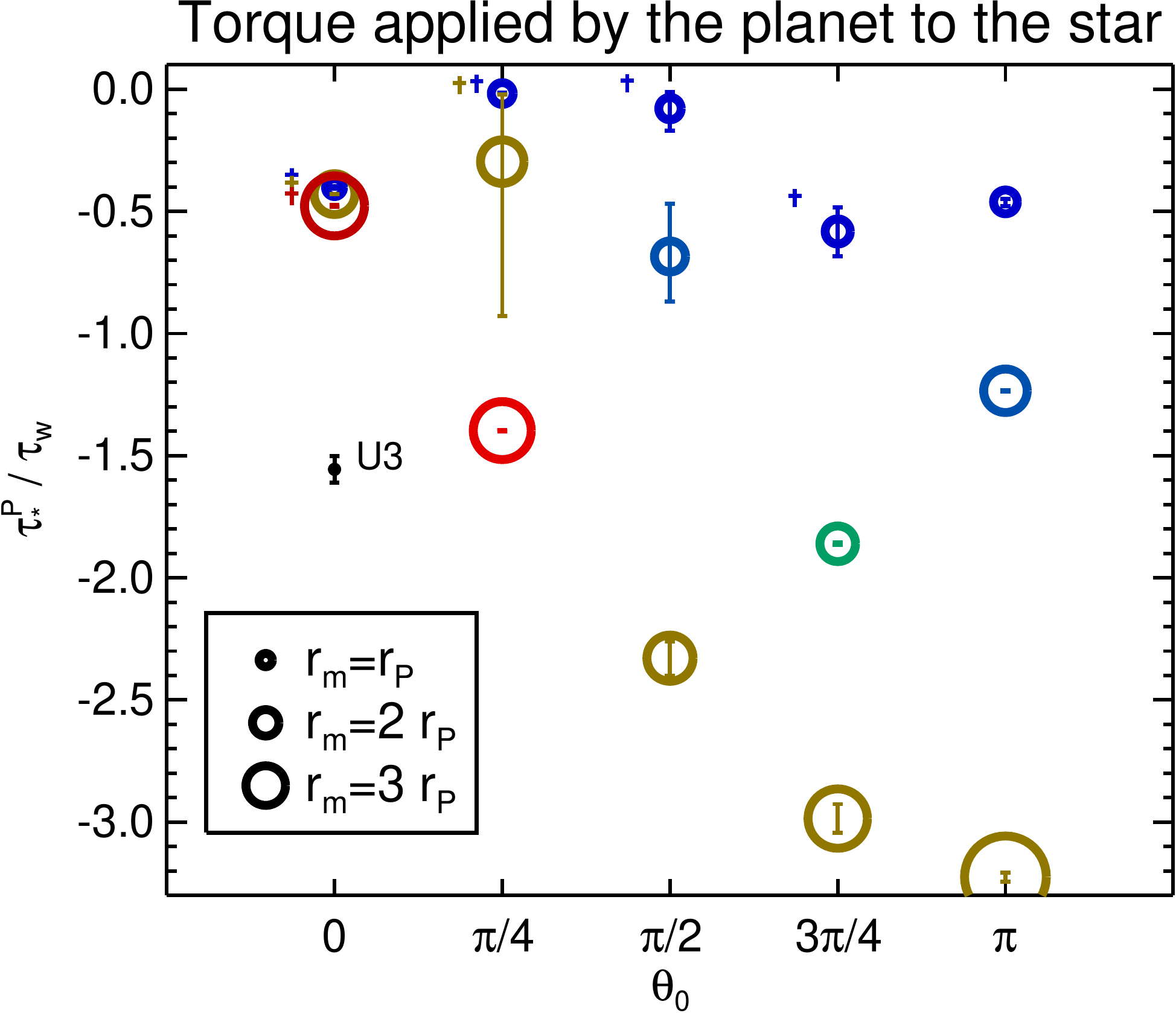}
    \includegraphics[width=\linewidth]{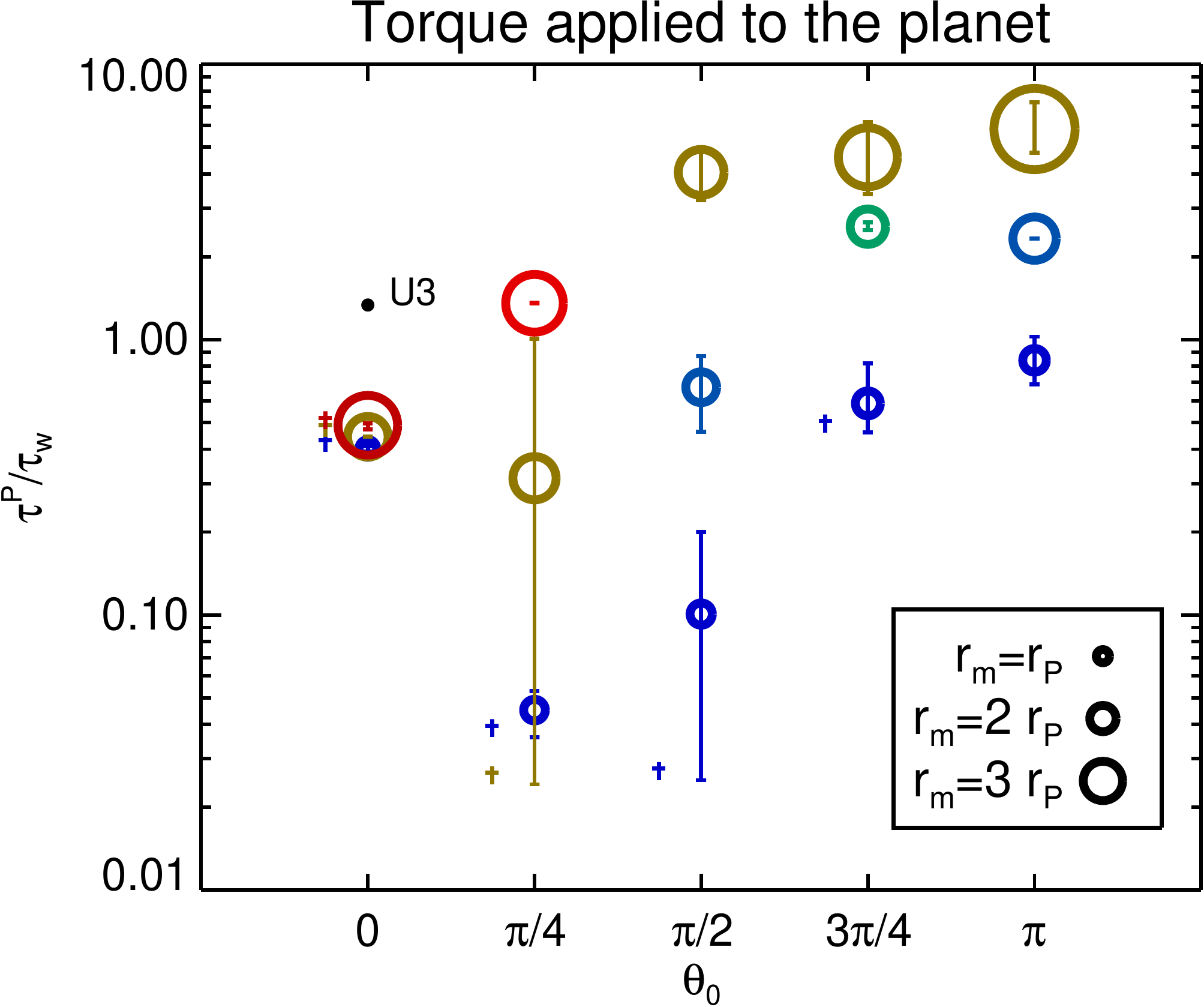}
  \caption{Torques for different planetary field amplitude (colored
    symbols, from \modif{blue--weak to green/yellow--medium to red--strong} magnetic fields) and
    different inclination angles (abscissa). The size
    of the symbols represent the size of the planetary
    magnetosphere (see table \ref{tab:tab3}). We display from top to bottom $\tau^{\star}_{w}$, $\tau^{\star}_{P}$ and $\tau^{P}$. In the lower two panels
    the closed configurations are labeled by \textdagger (see text). The unipolar
    case U3 is added for reference. Its error bars are obtained for
    variations over 4 orders of magnitude of the density level of the planet.}
  \label{fig:torq_magscan}
\end{figure}

The orientation and strength of the planetary magnetic field determines
the magnetic coupling efficiency between the two bodies (see Section
\ref{sec:dipolar-interaction}). The 
effect of the magnetic field orientation on its coupling with the coronal field
for a close-in planet was initially studied by \citet{Ip:2004ba}. In the present
work, we include the orbital motion of the planet, which
leads to the angular
momentum transfer. We
display in Figure \ref{fig:torq_magscan} the torque applied
by the wind (upper panel) and by the planet (middle panel) to the
star as a function of the inclination angle $\theta_{0}$ of the
planetary magnetosphere for a planet located at $r_{orb}=3\,r_{\star}$. All
torques are normalized to the fiducial stellar wind torque. The
averaged size of the planetary magnetosphere, measured at the magnetic
equator, is given by the size of the circles in the two panels (see
Table \ref{tab:tab2}). Their color labels the amplitude of the
planetary field, from the smallest (blue, $\mu_{P}=4.5\, 10^{-5}\mu_{\star}$) to
the largest (red , $\mu_{P}= 4.5\, 10^{-3} \mu_{\star}$). The
error bars are a measure of the time variations of the torques that occur either because
of a flipping between a closed and an open configuration, or because
of time-dependent magnetic reconnection on the nightside of the planet. For
each inclination angle, the wind modification and
$\tau^{\star}_{P}$ are directly correlated to the
strength of the planetary field (or magnetospheric size). However, the
magnetospheric size (or $B_{P}$) alone does not appear to be a good proxy to estimate the
strength of the SPMI \modif{because of the strong dependence on $\theta_{0}$}. The two extreme cases
D3a2 and D3r2 reveal the importance of considering the inclination
angle of the planetary fields: for comparable magnetospheric
dimensions, the torque applied to the planet (bottom panel of Figure
\ref{fig:torq_magscan}) varies by an order of magnitude. We
generally find that for a given magnetospheric size or for a given $\mu_{p}$, the torques
vary significantly with the inclination angle.

For a given magnetospheric size, the maximum stellar
wind modification (upper panel) is obtained for the aligned ($\theta_{0}=0$)
configuration. All the aligned cases are in the closed configuration.
As a result, for a given magnetospheric
size, the magnetic field lines connecting the star to the planet are
likely to extend further away from the ecliptic plane and impact
higher latitudes --where the wind is driven-- at the stellar
surface. This is particularly clear in Figure \ref{fig:fig_dipolar_example}
where it can be observed that the connecting 
field lines impact the stellar surface around $\theta_{i}\sim
47^{\circ}$ in the aligned case and around $\theta_{i}\sim 45^{\circ}$
in the anti-aligned case. 
Hence, the maximum wind modification
is likely to be obtained in the aligned topology. Conversely,
all the anti-aligned cases ($\theta_{0}=\pi$) are in the open
configuration. As a result, the magnetic link strengthens and is slightly
\modif{more concentrated} as the planetary
magnetic field increases. The interacting zone at the stellar surface
hence diminishes slightly, and the modification of the stellar wind
decreases when the amplitude of the planetary field is increased,
leading to an opposite behavior in the completely anti-aligned case,
compared to all other inclinations.

\begin{figure}[tbp]
    \includegraphics[width=\linewidth]{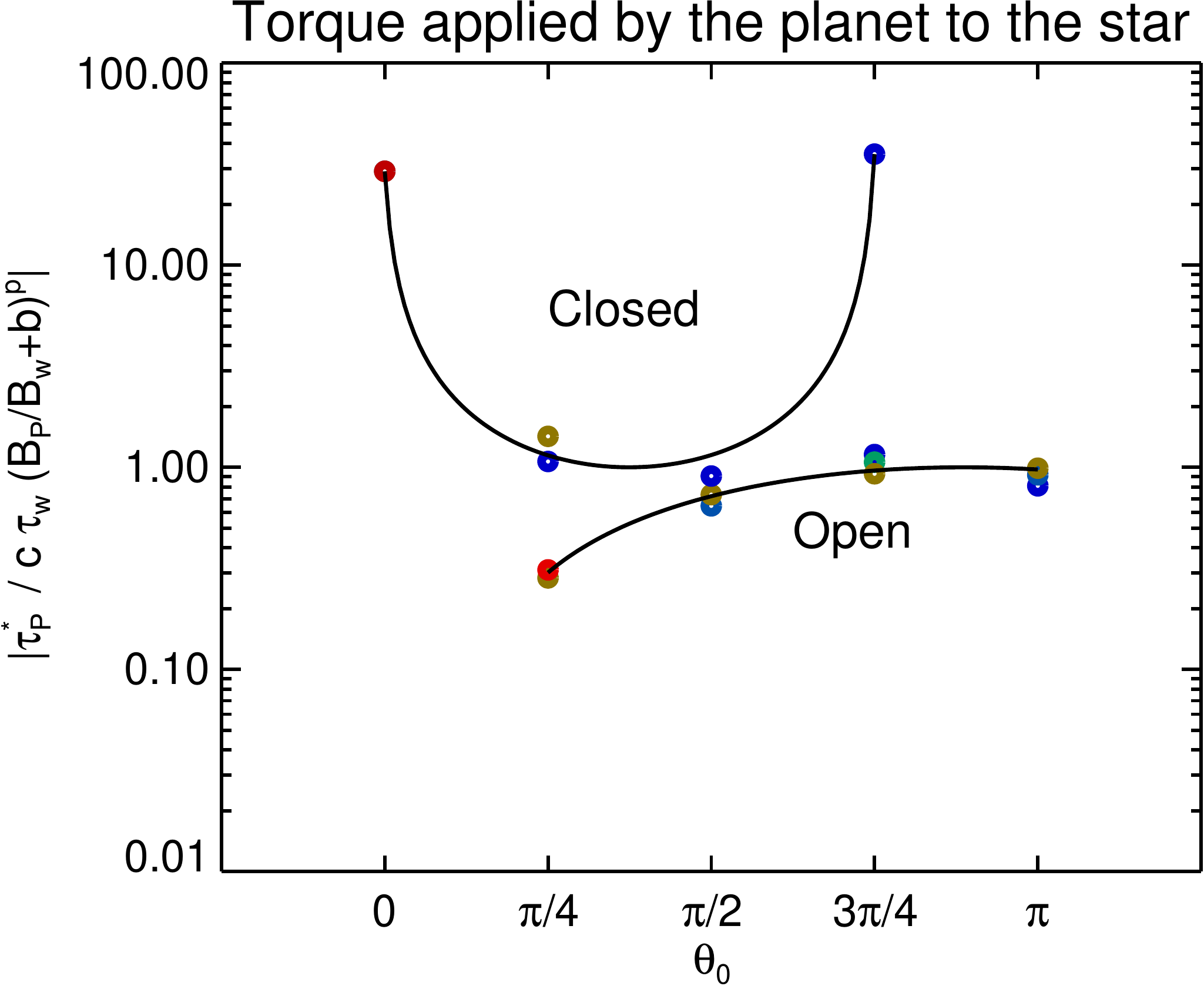}
    \includegraphics[width=\linewidth]{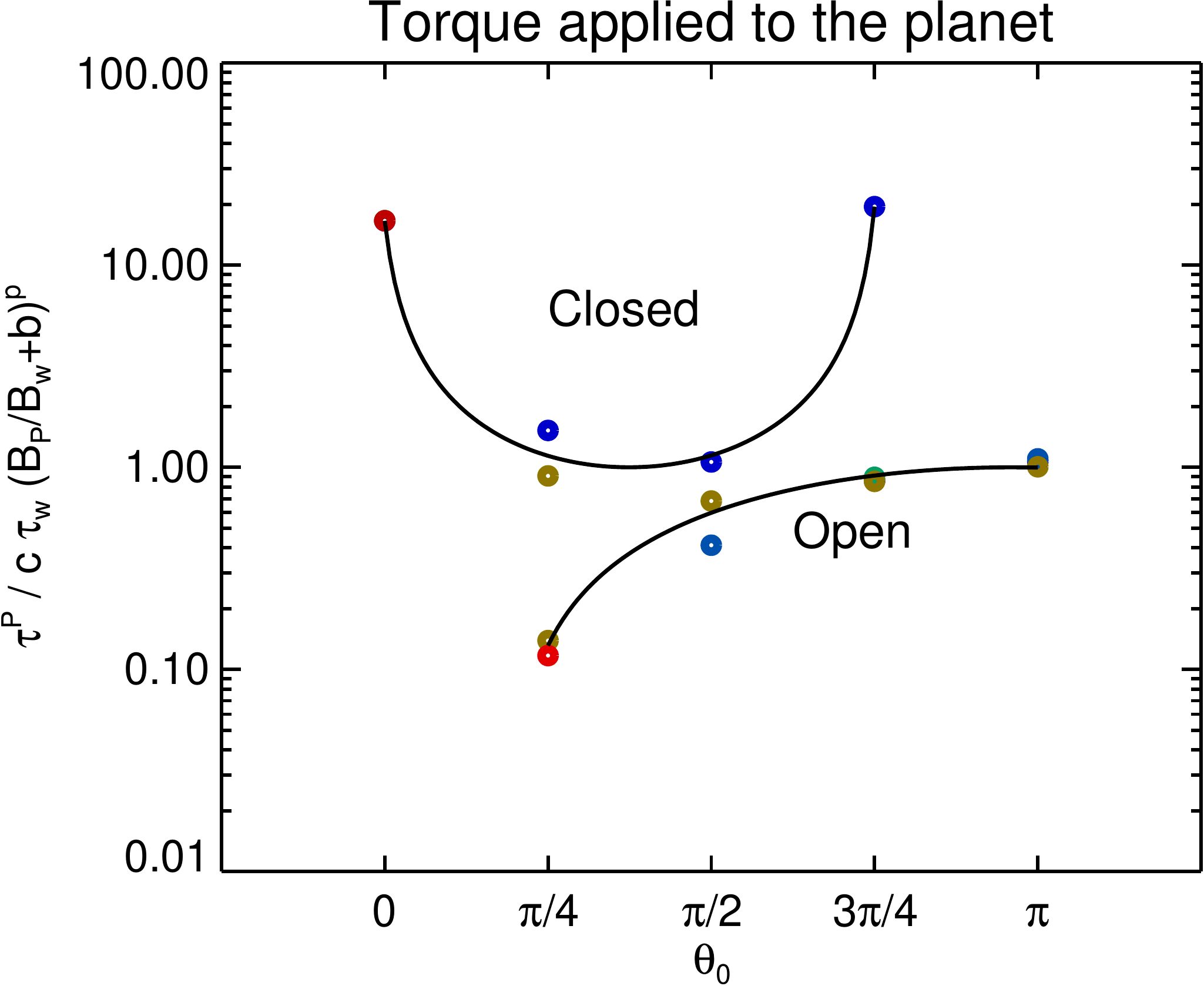}
  \caption{Fit (black lines) of the torque formulation \eqref{eq:taups} in the
    dipolar cases. The coefficients of the fits are given in Table \ref{tab:tab3}.}
  \label{fig:torq_magscan_fit}
\end{figure}

The torque applied by the planet to the star (middle panel) and the
total torque applied to the planet (lower panel) further reveal different
behaviors for the open and closed configurations.
In the closed configuration \modif{(cases labeled with daggers \textdagger)}, both torques
have a surprisingly very weak sensitivity to
the amplitude of the planetary field $B_{P}$ (note that in the
inclined cases switching configuration over time, the
closed configuration torques are
taken at the extremum of the error bars). The torques developing in
the open configuration, conversely, depend strongly on both $B_{P}$
and $\theta_{0}$. We fit $\tau^{\star}_{P}$ and
$\tau^{P}$ with the simple formulation
\begin{equation}
  \label{eq:taups}
  \frac{\tau}{\tau_{w}} = c
  \left(\frac{B_{P}}{B_{w}}+b\right)^{p}\cos^{t}\left(\frac{\theta_{0}-\Theta}{s}\right) 
\end{equation}
for each configuration (open 'o' and closed 'c'). \modif{We tested various formulations of the fitting
  function \eqref{eq:taups} until an acceptable fit was obtained for
  the various torques in both the closed and open configurations.} The fit
coefficients are shown in Table \ref{tab:tab3} and the fits are visualized in
Figure \ref{fig:torq_magscan_fit} (black lines). The slight
discrepancy from
the fits is a reasonable trade-off to provide the simple torque
formulation \eqref{eq:taups} for the dipolar interaction. Both torques
exhibit the same qualitative behavior. 

The magnetic interaction is minimized in
the closed configuration for an inclination angle close to
$\pi/3$. This results from a simple geometrical constrain in the
closed configuration. The field lines connecting the planet to the
star --the field lines that are responsible for the transfer of
angular momentum-- are necessarily anchored at the magnetic poles of
the planet. Each planetary pole has to be connected to one --and only
one-- of the stellar hemispheres in a stable magnetic configuration
(otherwise, strong currents would develop at the planetary poles and
the associated magnetic reconnections would tend to suppress such
configuration). As a result, in the aligned case the south
(resp. north) pole of the planet is connected to the north
(resp. south) hemisphere of the star. On one hand, if the planetary dipole is
slightly titled, the magnetic connection between the pole the most
distant from the star and the stellar hemisphere is harder to
establish and the overall magnetic link is weakened. On the other
hand, for a perpendicular dipole ($\theta_{0}=\pi/2$, middle panel in
Figure \ref{fig:fig_dipolar_example}), one of the planetary
pole is sufficiently close to the star to reverse this effect. As a
consequence, the effective torque is minimized for intermediate
inclination angles in the closed configuration. The exact
angle is likely to depend on the orbital radius or, said differently,
on the relative position of the planet inside the dead-zone. For a
planet located at $r_{orb}=3\,r_{\star}$ we found that the torque is
minimized for $\theta_{0}\sim 0.37\pi$.

In the open configuration the torques are maximized for
$\theta_{0}\sim \pi$. The anti-aligned ($\theta_{0}=\pi$) planetary
field is naturally compatible with the dipolar structure of the
stellar magnetic field, with vertical (along the $z$ direction) field
lines near the ecliptic. When the planetary field is inclined from this
configuration, the magnetic links at the two poles of the planet are
likely to shrink to accomodate the topological constraint. As a result,
the connection between the poles of the planet and the
stellar hemispheres are weaker and the associated torques
decrease, as observed in Figure \ref{fig:torq_magscan_fit}.


\begin{deluxetable}{lcccccc}
  \tablecaption{Torque coefficients in the dipolar case\label{tab:tab3}}
  \tablecomments{Closed (c) and open (o) configuration are
    distinguished. The fit coefficients are defined in Equation
    \eqref{eq:taups}.
}
  \tablecolumns{7}
  \tabletypesize{\scriptsize}
  \tablehead{
    \colhead{$\tau$ $[\tau_{w}]$} &
    \colhead{$c$} &
    \colhead{$b$} &
    \colhead{$p$} & 
    \colhead{$t$} &
    \colhead{$\Theta$} &
    \colhead{$s$}
  }
  \startdata 
  $\tau^{\star}_{w}$ (c)  & 4.13  & 492  & -0.24 &
  1 & 0.45 $\pi$ & 2.08 \\
  \hline\\
  $\tau^{\star}_{P}$ (c)  & -0.001 & 1140  & 0.34  &
  -1 & 0.37 $\pi$ & 0.77 \\
  $\tau^{\star}_{P}$ (o)  & -0.27 & -27.7 & 0.43  &
  1 & 0.89 $\pi$ & 1.58 \\
  \hline\\
  $\tau^{P}$    (c)  & 0.02  & 37.0  & 0.07  &
  -1 & 0.37 $\pi$ & 0.78 \\
  $\tau^{P}$    (o)  & 0.17 & -19.54 & 0.65  &
  1 & 0.96 $\pi$ & 1.55 
  \enddata
\end{deluxetable}

\subsection{Orbital radius dependence}
\label{sec:unip-dipol-inter}

\begin{figure*}[tbp]
    \includegraphics[width=0.45\linewidth]{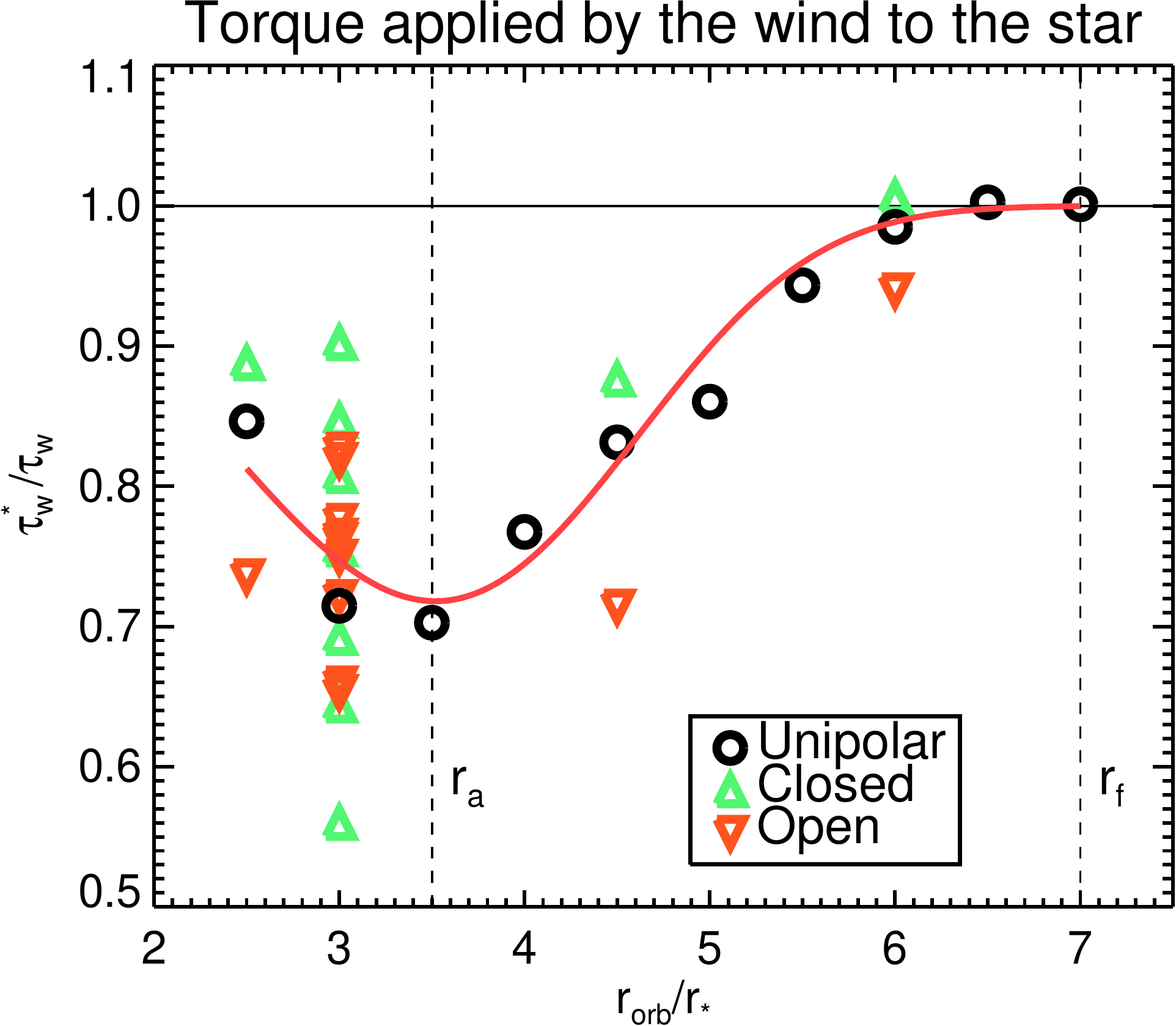}
    \includegraphics[width=0.45\linewidth]{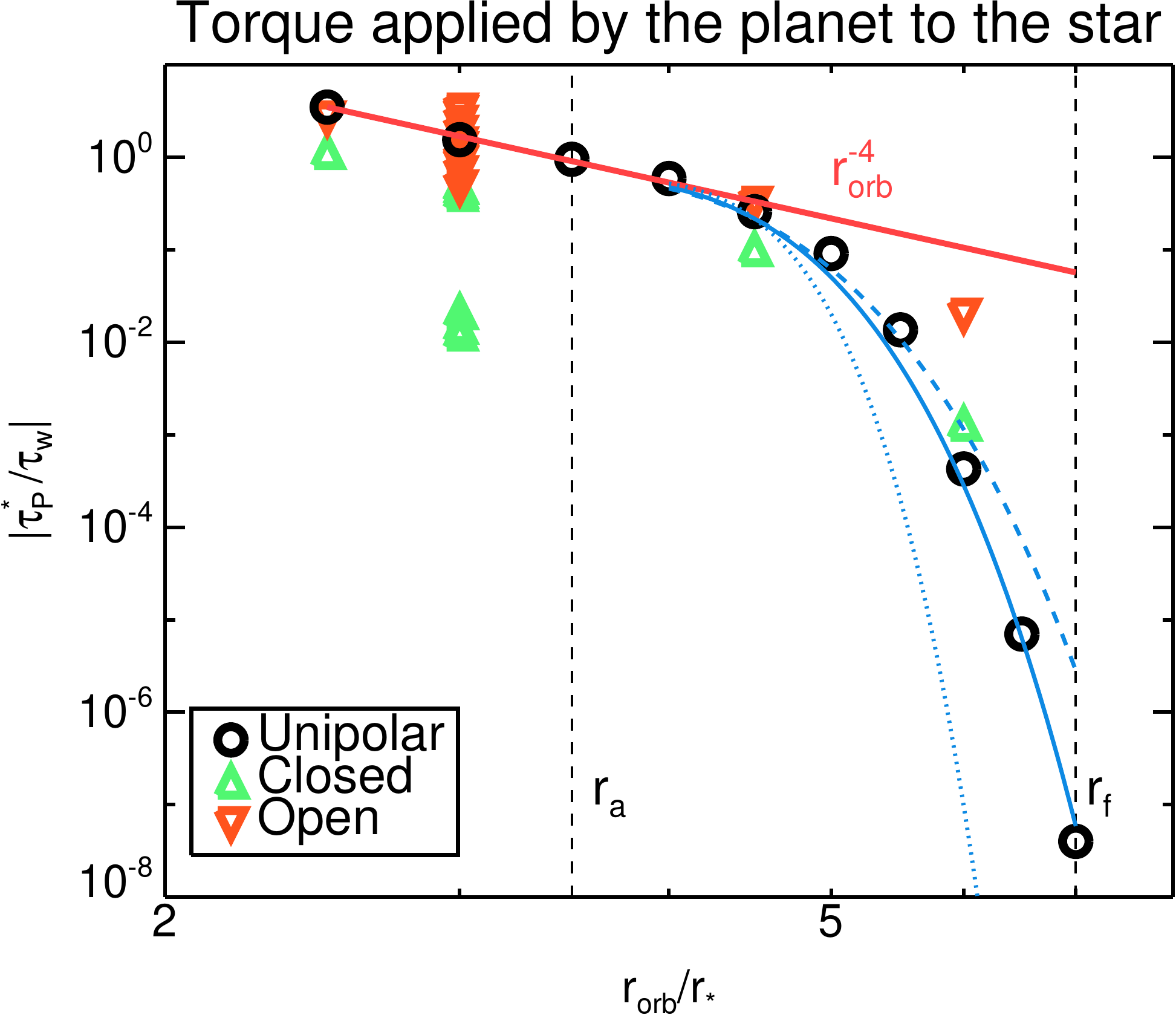}
  \caption{Normalized magnetic torques as a function of the orbital
    radius. Black circles label the unipolar cases, and
    the upward and downward triangles the dipolar cases in the closed
    and open configurations. 
    The left panel shows the torque applied by the 
    stellar wind. The modification of the stellar wind
    torque is fitted with a Weibull distribution (red curve, see text). The right panels
  shows the torque applied by the planet to the star with logarithmic
  scales. The cases with a planet inside the Alfv\'en radius $r_{a}$
  are well matched by a $r^{-4}_{orb}$ power law. In the two panels
  the position of the Alf\'ven $r_{a}$ and fast Alv\'en $r_{f}$
  surfaces on the equator are shown by the vertical dashed lines.}
  \label{fig:torq_radscan}
\end{figure*}

\reff{Up to} now we have \modif{discussed cases with} an orbital radius of $3\,
r_{\star}$. We now \modif{present cases with variations} from $2.5\, r_{\star}$ to $7\,
r_{\star}$ --spanning the whole \modif{sub-alfv\'enic} zone-- for the
unipolar interaction case. We also ran some dipolar cases for particular
orbital radii (see
Table \ref{tab:tab2}). We
display the resulting wind and planetary 
torques as a function of $r_{orb}$ in Figure \ref{fig:torq_radscan}. 

The stellar wind torque is systematically reduced by the orbiting
planet (left panel). The decrease is maximized for planets orbiting close to
the \modif{streamer boundary (the transition region between open and
  closed field lines of the stellar wind)}. The magnetic connection has a
tendency to inhibit the wind driving at its foot-point on the stellar
surface. If the planet orbits too close to the star, \modif{the foot-point of
the magnetic link on the stellar surface} is
located well inside the closed field lines region where only a slow
(or even no) wind is driven: the SPMI does not significantly modify the stellar
wind in this
case. If the planet orbits in the open field lines region --and
inside the fast Alfv\'en radius--, the foot-point latitude
does not change much with the orbital radius, since the planet remains
on the equatorial plane. As a consequence, the strength of the SPMI
decreases with the orbital radius due to the combined effects of the
decrease of $\left|v_{P}-v_{\phi}\right|$ ($v_{P}\propto
r_{orb}^{-1/2}$) and the decrease of the stellar wind magnetic
field. We match the modification of the stellar wind torque in the
unipolar case
$\Delta\tau=1-\tau^{\star}_{w}/\tau_{w}$ with a Weibull
distribution (red line in the left panel of Figure
\ref{fig:torq_radscan}) that is given by
\begin{equation}
  \label{eq:weibull}
  \Delta \tau  = \Delta \tau_{0}
  \left(\frac{r_{orb}}{r_{i}}\right)^{k-1}e^{-\left(\frac{r_{orb}}{r_{i}}\right)^{k}}\, ,
\end{equation}
with $(\Delta \tau_{0},r_{i},k)$ being the free parameters for the
fit. We find that the location at which the modification is maximized at the node of the
Weibull distribution $r_{i}((k-1)/k)^{1/k}\sim 3.5\, r_{\star} = r_{a}$.
The power-law exponent is found to be $k \sim 3.9$. 

The torque applied by the planet to the star (right panel with
logarithmic axes) also shows two
different trends inside and outside the Alfv\'en radius $r_{a}$. For
planets orbiting inside $r_{a}$, the torque decreases like
$r_{orb}^{-4}$. The only
other estimation we are aware of was provided
by \citet{Laine:2011jt} in the unipolar case. They found (see their
Equation 15, adapted for
circular orbits) that the torque scales like
$r_{orb}^{-5.5}$. The discrepancy may originate from two
effects. First, the 2.5D geometry we consider 
over-estimates the magnetic torques. One can crudely extrapolate our results
by rescaling the torques with a geometrical factor $\alpha_{g}=r_{P}/\pi r_{orb}$,
which would make the torque decrease like
$r_{orb}^{-5}$. The remaining power-law difference with the work of
\citet{Laine:2011jt} could be due to a more subtle geometrical
effect. However, it is also likely that it arises from the fact that
we consider a magnetic field self-consistently evolving
with the dynamical wind, and not constrained to be purely dipolar.

If the planet is orbiting in between the Alfv\'en and the
fast-Alfv\'en radius, the torque it applies to the star falls \modif{off}
exponentially. The planet in this case is
exposed to the \modif{accelerating} wind, making
it hard to establish a magnetic connection with its host. We fit this
fall off in the unipolar case with the function (solid blue
line in Figure \ref{fig:torq_radscan})
\begin{equation}
  \label{eq:fit_exp_falloff}
  \tau^{\star}_{P} =
  \tau^{\star}_{P}\left(r_{a}\right)
  e^{-(r_{orb}-r_{a})^{k_{f}}}\, ,
\end{equation}
with $k_{f}=2.3$. Note that the exact exponential
fall-off may differ between the dipolar and unipolar cases. For
reference we plotted the same curve for
$k_{f}=2$ (dashed blue line) and $k_{f}=3$ (dotted blue line). 

\modif{The dipolar cases (orange and green triangles in the two panels
  of Figure \ref{fig:torq_radscan}) show a
  significant spread across the scaling law we derived from the unipolar
  cases. By using Equation \eqref{eq:taups} to get rid of the
  $(B_{P},\theta_{0})$ dependence in the closed and open
  configurations, this spread is significantly reduced (not shown here).
  The dipolar cases are then observed to follow the
  same trends with the orbital radius, albeit with different
  proportionality factors that could be properly determined with
  a larger set of dipolar models at different orbital radii.
}

It was previously argued that a SPMI involving a close-in planet would
generally decrease the stellar wind angular momentum loss because the
planet would block a significant part of the outgoing stellar wind
\citep{Cohen:2010jm}. Here we also find --provided the star rotates sufficiently
slowly, see Figure \ref{fig:fig_schematic}-- that SPMI generally reduces
the magnetic torque applied to the central star. \modif{We
  find that the net torque applied
  to the star is decreased primarily because of the direct transfer of
  angular momentum from the planet to the star. The SPMI also leads to
  a modification of the wind driving at the stellar surface, which in
  turns leads to a reduction of the
  torque applied by the wind to the star. This effect, which we
  believe to be at the origin of the 'blocking' effect identified by
  \citep{Cohen:2010jm}, is found to be generally less important.
Our results nevertheless also support the idea that SPMI}
could partly explain the empirical evidence of excess of
rotation observed in stars hosting close-in planets \citep[see also][]{Pont:2009ip}.

\subsection{Planet migration}
\label{sec:planet-migration}

\reff{The magnetic link that connects the planet and the star together 
leads to a torque that applies to the two celestial bodies. The resulting angular
momentum transfer changes the stellar rotation and
the planetary orbit. The spin-angular momentum $J$ of
the host star and the orbital angular momentum of the planet can be defined by
\begin{eqnarray}
  \label{eq:mom_inertia1}
  J_{\star} &=& I_{\star} \Omega_{\star} \sim k^{2} M_{\star} r_{\star}^{2}
  \Omega_{\star} \, ,\\
  \label{eq:mom_inertia2}
  J_{P} &=& I_{P} \omega_{orb} = M_{P} r_{orb}^{2} \omega_{orb}\, ,
\end{eqnarray}
where $I$ represents the moment of inertia and $\omega_{orb} =
\sqrt{GM_{\star}/r_{orb}^{3}}$. \modif{The normalized radius of gyration $k^{2}$ is
  of order $0.1$ for a main sequence solar-like star.} $I_{\star}$ is
generally slightly higher than $I_{P}$ for close-in planets.
The angular momentum ratio is given by
\begin{equation}
  \label{eq:amom_ratio}
  \frac{J_{P}}{J_{\star}} = \frac{M_{P}}{k^{2}M_{\star}}
\left(\frac{r_{orb}}{r_{\star}}\right)^{1/2} f^{-1}\, .
\end{equation}
\modif{The full range of our models is between $J_{P}/J_{\star}\sim 40$ (for $r_{orb}=3\,r_{\star}$) and
$60$ (for $r_{orb}=7\, r_{\star}$)}, which shows that, in all cases,
the orbital angular momentum of the planet is higher
than the rotational angular momentum of the star. This dominance is due to the
  relatively slow rotation of the star (the planet, in this work, is always inside
  the co-rotation radius, see Figure \ref{fig:fig_schematic}).}

\begin{figure}[tbp]
  \includegraphics[width=\linewidth]{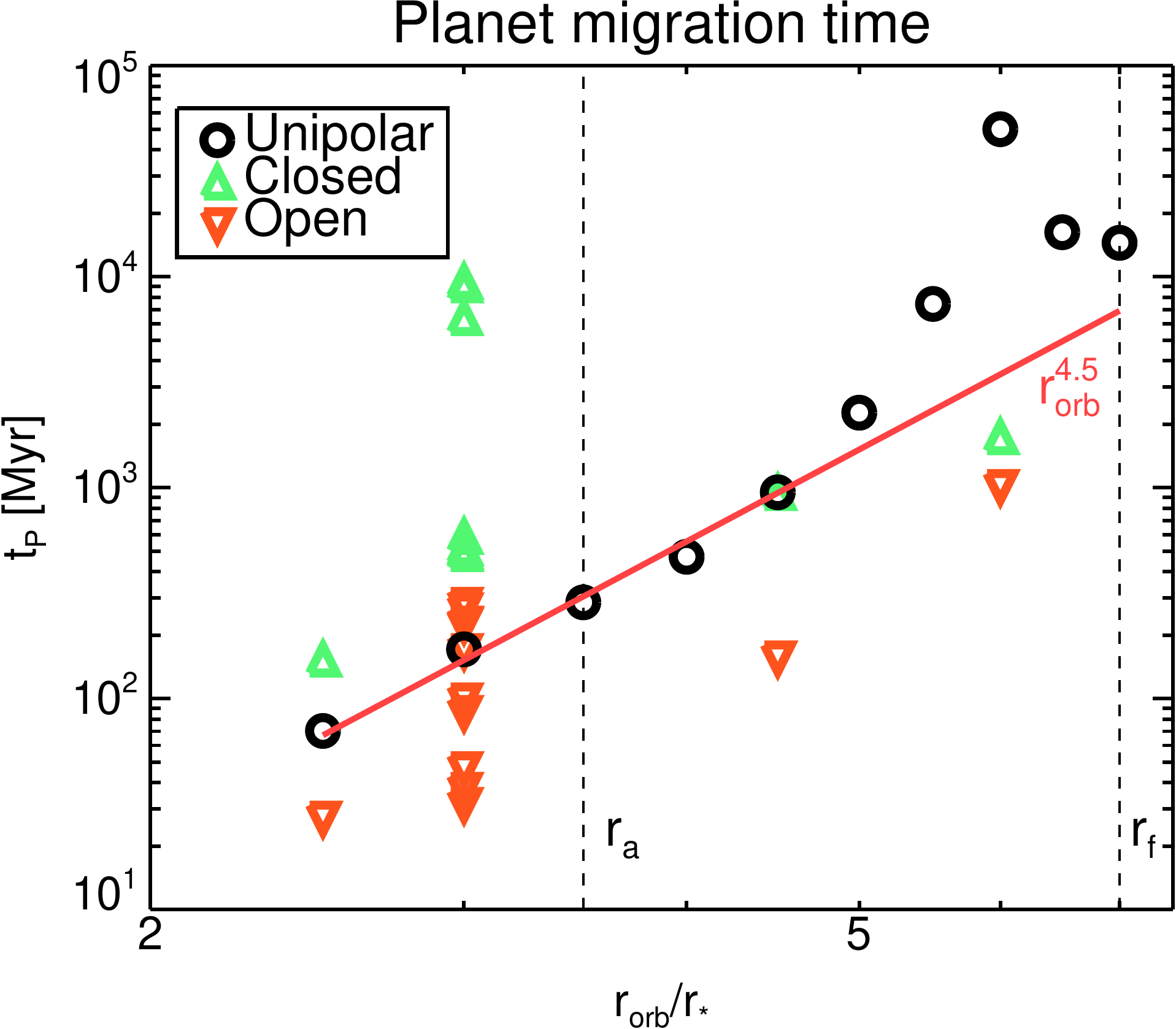}
  \caption{Migration time scale as a function of the orbital radius
    for $B_{\star}=246$ G. The migration time-scale is fitted by a
    power-law, shown in red (see text for details).}
  \label{fig:migtime_p}
\end{figure}

\reff{Whether or not the effect of SPMI is significant
  depends on the time-scale over which angular momentum is transfered
  between the two bodies.
Based on the angular momentum definitions
\eqref{eq:mom_inertia2}, we define the
evolution time-scale
\begin{eqnarray}
  \label{eq:ts_planet}
  t_{P} &=& \frac{r_{orb}}{\dot{r}_{orb}} =
  2\frac{J_{P}}{|\tau^{P}|} \, ,
\end{eqnarray}
where the factor $2$ comes from the
$r_{orb}^{1/2}$ dependence of $J_{P}$.
We plot the migration time-scales assuming the case of a large stellar
magnetic field, $B_{\star}=246$ G
in Figure \ref{fig:migtime_p}. 
In the unipolar case and dipolar cases in the closed
configuration, the amplitude of the torque applied to the star by the planet
and of the net torque applied to the planet are very similar,
since the planet loses a negligible amount of angular momentum to
the wind. Based on the scaling found in Section
\ref{sec:unip-dipol-inter}, we expect the migration time to scale
with $r_{orb}^{4.5}$ (see Equation \ref{eq:ts_planet}). We
overplot this scaling in red in Figure \ref{fig:migtime_p} and
indeed observe a general agreement for the innermost planets.
}

\modif{The migration time in the unipolar case then increases
  exponentially in between the Alfv\'en surfaces, but appears to go
  back to a constant value near the fast Alfv\'en
surface boundary. In this region the planet is subject to a direct interaction with the
fast stellar wind and hence starts --in the unipolar case-- to lose angular momentum to the
wind only. The dead-zone rotates at the stellar rotation rate whereas the
open field line region trails behind at a lower rotation
rate. When the planet is completely outside the dead-zone, it interacts with a more
slowly rotating plasma which enhances the differential motion driving
the SPMI. As a result, we observe a discontinuity in the migration
time-scale of the planet. It is worth noting that, even if the planet
were outside the fast Alfv\'en radius, it would still undergo an
orbital decay linked to its interactions with the rotating wind. It
would eventually enter the dead-zone and magnetically interact with
its host\modif{, if the system lives long enough}. We ran a couple of
other models (not shown here) with orbital radii between $7$ and $10\,
r_{\star}$ showing that the decay time-scale remains approximately constant
within a few stellar radii outside the Alfv\'en surface.
}

\modif{The migration time-scales in the dipolar cases span two orders of
magnitude, depending on the amplitude and inclination of the planetary
magnetic field. We note again that the topology of the planetary field
has a dramatic influence on the SPMI: a planet in the open
configuration will systematically lose orbital angular momentum much faster than a planet in
the closed configuration. For more realistic
stellar magnetic configurations, the topology is likely to
switch back and forth aligned and anti-aligned configurations as the
planet orbits inside the complex coronal magnetic field.
Our results show that the time-averaged torque applied to the
planet would be largely dominated by anti-aligned phases. Hence, such configuration
should be used when estimating migration time-scales due to SPMIs
in real star-planet systems.
\reff{The dipolar cases in the open configuration seem to roughly follow the 
same power-law as the
unipolar and closed cases, for fixed values of inclination and strength of
the planetary field (not shown here). However}, the grid of
models presented here does not allow us to properly verify 
the power-law exponent in the open configuration, which
would require additional simulations at several orbital radii. In
addition, it must be stated that a planetary outflow \reff{(powered
by, \textit{e.g.}, stellar
radiation)} could also modify this power-law.}

\reff{The planet migration-time scales with
  $B_{\star}^2$. If one considers a star with a lower magnetic field
  (\textit{e.g.} $B_{\star}=0.78$ G), the migration time-scales shown in figure
  \ref{fig:migtime_p} are five orders of magnitude larger, and the SPMI
  clearly does not play a significant role in the planet migration
  during the whole system life-time. On the other hand, for the larger
  stellar magnetic field shown ($B_{\star}=246$ G), the planet
  migrates inside the dead-zone on a time-scale ranging from approximately
  $20$ to $20000$ Myr. Thus, in some cases the
  times-scales can be mush shorter than the secular evolution of the
  system: in these cases SPMI is
  certainly one of the major effects on the planetary orbit.}

\section{Toward general formulations of the torques}
\label{sec:discussion}

In order to compile all the SPMI effects together for the unipolar and
dipolar interactions, we propose hereafter a
formulation for the torques applied to the star and the
planet. \reff{We also apply} the geometrical factor $\alpha_{g}$
to account for the reduced 2.5D geometry used in this work. 

The final torque that applies to the star results from the modified
stellar wind and from the torque applied to the planet. It can be
written in the following form
\begin{equation}
  \label{eq:torq_star}
  \tau^{\star} = \tau_{w}\left\{
    Q_{\Upsilon} + Q_{P} \right\} \, ,
\end{equation}
where $\tau_{w}$ is the fiducial wind torque which can be obtained from \citet{Matt:2012ib,Reville:2014ud},
$Q_{\Upsilon}$ is the ratio quantifying the modification of the open flux
due to the SPMI, and $Q_{p}$ is the normalized torque associated
with the orbiting planet. From Section \ref{sec:magnetic-torques} we
can write 
\begin{equation}
  \label{eq:qupsilon}
  Q_{\Upsilon} = 1 - \alpha_{g}\Delta\tau\, ,
\end{equation}
with $\Delta\tau$ given by Equation \eqref{eq:weibull}. $Q_{P}$ is
given by
\begin{equation}
  \label{eq:qplanet}
  Q_{P} = \left(\frac{r_{orb}}{r_{\star}}\right)^{-5}
  \mathcal{H}_{P}\, ,
\end{equation}
where $\mathcal{H}$ is defined in the
unipolar and dipolar cases by
\begin{eqnarray}
  \label{eq:hplanet_u}
  \mathcal{H}^{u}_{P}  &=& C_{0} \, \\
  \label{eq:hplanet_p}
  \mathcal{H}^{d}_{P} &=&
  C_{1} \left(\frac{B_{P}}{B_{w}}+b\right)^{p}\cos^{t}\left(\frac{\theta_{0}-\Theta}{s}\right)\, .
\end{eqnarray}
The parameters defining $\mathcal{H}^{d}_{P}$ can be found in Table
\ref{tab:tab3} for both the open and closed configurations.

The total magnetic torque that applies to the orbiting planet includes
the torque from the star and the one from the wind. We combine the
results from Sections \ref{sec:incl-plan-magn} and
\ref{sec:unip-dipol-inter} to obtain the following torque formulation
\begin{equation}
  \label{eq:overall_torq_p}
  \tau^{P} = \tau_{w} C_{2} \left(\frac{r_{orb}}{r_{\star}}\right)^{-5}
  \left(\frac{B_{P}}{B_{w}}+b\right)^{p}\cos^{t}\left(\frac{\theta_{0}-\Theta}{s}\right)\, 
\end{equation}
where the parameters $(b,p,t,\Theta,s)$ depend on the configuration and are given
in Table \ref{tab:tab3}. 

The exact multiplicative constants  ($C_{0}$, $C_{1}$ and $C_{2}$) are likely
to depend on the radius of the
planet, which we did not vary in this study. $C_{0}$ also
depend on the grid resolution and should also be calibrated with simulations
done in 3D geometry. Hence, the
numerical value of those multiplicative constants needs to be
considered with caution. 

\section{Conclusions}
\label{sec:conclusions}

In this work we have explored the efficiency of the magnetic
interactions between a star and a close-in planet
to transfer angular momentum. We
explored the differences between the cases of a magnetized (dipolar
interaction) and non-magnetized (unipolar interaction) planets. Our
results can be summarized as follows.
\begin{itemize}
\item[-] The SPMI systematically decreases the torque applied by the stellar
  wind. This effect is maximized when the planet is in orbit close to
  \modif{streamer boundaries} (the
  open/closed field lines transition region) in the corona.
\item[-] When the star rotates slowly (as considered here), the torque
  applied by the planet to the star is
  generally higher than the decrease of the torque applied by the
  wind. It can even compete with the total angular momentum removed by the
  wind, and in some cases result in a
  net increase of angular momentum for the star.
\item[-] The torque applied by the planet to its host star
  is qualitatively similar in the unipolar and in the dipolar
  cases, but differs significantly \modif{in its amplitude}.
\item[-] Two magnetic configurations can be encountered in the dipolar
  case, where the planetary magnetosphere is either confined around
  the planet (the so-called closed configuration) or where it opens in
  the stellar \reff{wind} (the so-called open configuration).
\item[-] In the dipolar case, the angle of inclination of the
  planetary field with respect to the coronal field
  can greatly modify the efficiency of the SPMI. \modif{The knowledge of both
  the magnetospheric size and the inclination angle of the planetary
  field is \reff{needed} to estimate the angular momentum transfers between the
  star and the planet.}
\item[-] The planet migration associated to the SPMI
  is unstable \modif{inside the Alfv\'en surface}. A planet in orbit
  inside (resp. outside) the co-rotation
  radius will systematically migrate inward (resp. outward). Furthermore, the magnetic
  interaction with a planet inside the Alfv\'en surface strengthens with time and
  leads to an accelerated decay of the orbiting
  planet. \reff{This effect may be counterbalanced
  only if other processes (\textit{e.g.}, tidal forces in multi-planet
  systems)
  are taken into account}. \modif{Provided the star is rotating sufficiently
  fast, a stable point for the SPMI can also exist outside the
  Alfv\'en surface where the planet is in
  co-rotation with the rotating wind.}
\item[-] \modif{The migration time-scale linked to the SPMI is shown to be
  sufficiently short in some cases (particularly when the magnetic
  fields are strong)\reff{, demonstrating the SPMI} to be a first order effect in the
  secular evolution of the
  star-planet system (see Section \ref{sec:planet-migration}).}
\item[-] \modif{The torque applied to the planet, in the dipolar interaction case, is strongest
    in the open configuration. Therefore the open configuration
    \reff{state is likely to dominate long-term transfers of angular
      momentum and }should
    be preferred to
    estimate the potential contribution of the SPMI to the
    planetary migration.}
\item[-] \reff{Empirical} scaling laws for the wind modification, the torque between
  the star and the planet, and the planet migration time were proposed and
  summarized in Section \ref{sec:discussion}.
\end{itemize}

\reff{To further refine the scaling laws we derived, several improvements
are needed}. The first obvious
limitation of this work lies in its reduced geometry. Fully 3D
simulations will be needed to adequately validate the scaling laws
(Section \ref{sec:discussion}), and especially the multiplicative
constants in front them (Equations
\ref{eq:weibull} and \ref{eq:hplanet_u}-\ref{eq:overall_torq_p}). They
would also allow us to take into account the eventual
rotation of the planet for non-synchronized star-planet systems. \reff{All
the types of unipolar interactions} that we did not considered here
(see Section \ref{sec:unipolar-interaction})
could also be explored with 3D simulations.

Second, real stars possess much more complex
magnetic structures in their corona. Even though our formulae
give first order estimates of the torques acting in the system, they
were derived with a dipolar topology of the stellar magnetic field,
\modif{for which the dead-zone is on the ecliptic. This is
  not the case for, \textit{e.g.}, a quadrupolar-dominated wind
  \citep[see, \textit{e.g.}][for wind simulations with various realistic magnetic
  topologies]{Pinto:2011ca}. In
  addition, real stellar magnetic field are generally non-axisymmetric. Hence,
calculating the torque in the general case would require 
3D simulations, to take into account their temporal and spatial variation
in the stellar corona obtained either from numerical simulations
\citep{Brun:2004ji,Ghizaru:2010im,Brown:2010cn,Kapyla:2012dg} or
observations \citep[see, \textit{e.g.},][]{Petit:2008aa,Fares:2013aa}.}

Third, we explored the SPMIs for a
planet with a fixed radius and a fixed mass. The dependence of the scaling
laws we proposed should also be characterized with respect to those
planetary parameters. 

Finally, we focused this first study on
magnetic interactions. It is certainly possible to
retain more physical effects in our model. Radiative transfer should
be taken into account, to be able to model the 'induced magnetosphere'
unipolar case (Venus-like) and planetary outflows
\citep{Trammell:2014aa}. Tidal interactions provide another \reff{major
mechanism for} angular momentum transfer and should
be treated self-consistently with the SPMI, in order to develop a
unified theory of angular momentum transfers in star-planet
systems. The model we presented in this work can be
used as a basis for a global modelling framework of
star-planet systems.

\acknowledgments

\reff{We thank an anonymous referee for valuable comments on our manuscript.}
AS thanks A. Cumming, R. Pinto, A. Vidotto and P. Zarka for 
discussions about stellar winds and star-planet interactions. This
work was supported by
the ANR 2011 Blanc 
\href{http://ipag.osug.fr/Anr\_Toupies/}{Toupies}
and the ERC project \href{http://www.stars2.eu/}{STARS2}. AS
acknowledges support from the Canada’s Natural Sciences and
Engineering Research Council. We acknowledge access to supercomputers
through GENCI (project 1623), Prace and ComputeCanada infrastructures.

\bibliographystyle{yahapj}
\bibliography{mybib}

\end{document}